\documentclass[12pt]{article}
\pdfoutput=1
\usepackage{graphicx} 
\usepackage{bm}
\usepackage{booktabs}
\usepackage{graphicx}
\usepackage{subcaption}
\usepackage{epstopdf}
\usepackage{epsfig}
\usepackage{bm}
\usepackage{xcolor} 
\usepackage{colortbl}
\usepackage{array}
\usepackage{multirow}
\usepackage{threeparttable}
\usepackage{amssymb}
\usepackage{amsthm,amsmath,amssymb}
\usepackage{natbib}
\usepackage{mathrsfs}
\usepackage{xr}
\usepackage{enumerate}
\usepackage{url}
\newtheorem{theorem}{Theorem}
\newtheorem{assumption}{Assumption}

\newtheorem{remark}{Remark}

\newtheorem{corollary}{Corollary}
\newtheorem{proposition}{Proposition}
	
\definecolor{Ocean}{RGB}{129,194,234}

\DeclareMathOperator*{\argmin}{argmin}

\DeclareMathOperator*{\E}{E}

\DeclareMathOperator*{\diag}{diag}

\newcommand{\blind}{1}

\begin{document}

\def\spacingset#1{\renewcommand{\baselinestretch}%
{#1}\small\normalsize} \spacingset{1}


\if1\blind
{
  \title{\bf Generalized Matrix Factor Model}
  \author{Xinbing Kong \thanks{School of Statistics and Data Science, Nanjing Audit University, Nanjing 211815, China. The author acknowledges the support from the NSF China (72342019, 12431009). Email: xinbingkong@126.com}
   \ \ and
    Tong Zhang \thanks{School of Statistics and Data Science, Nanjing Audit University, Nanjing 211815, China. Co-first author. Email: zhangtongsta@163.com}
}
  \maketitle
} \fi

\if0\blind
{
  \bigskip
  \bigskip
  \bigskip
  \begin{center}
    {\LARGE\bf Generalized Matrix Factor Model}
\end{center}
  \medskip
} \fi

\bigskip
\begin{abstract}
This article introduces a nonlinear generalized matrix factor model (GMFM) that allows for mixed-type variables, extending the scope of linear matrix factor models (LMFM) that are so far limited to handling continuous variables. We introduce a novel augmented Lagrange multiplier method, equivalent to the constraint maximum likelihood estimation, and carefully tailored to be locally concave around the true factor and loading parameters. This statistically guarantees the local convexity of the negative Hessian matrix around the true parameters of the factors and loadings, which is nontrivial in the matrix factor modeling and leads to feasible central limit theorems of the estimated factors and loadings. We also theoretically establish the convergence rates of the estimated factor and loading matrices for the GMFM under general conditions that allow for correlations across samples, rows, and columns. Moreover, we provide a model selection criterion to determine the numbers of row and column factors consistently. To numerically compute the constraint maximum likelihood estimator, we provide two algorithms: two-stage alternating maximization and minorization maximization. Extensive simulation studies demonstrate GMFM's superiority in handling discrete and mixed-type variables. An empirical data analysis of the company's operating performance shows that GMFM does clustering and reconstruction well in the presence of discontinuous entries in the data matrix.
\end{abstract}

\noindent%
\vfill

\newpage
\spacingset{1.9} 
\section{Introduction}

Matrix sequences appear in diverse areas, such as computer vision, recommending systems, social networks, economics, and management. It has the form of a three-way array consisting of a row dimension, a column dimension, and a third sequence limit dimension that might be a time domain. A natural question is ``how are the entries generated?" In the high-dimensional scenario where both the numbers of rows and columns are large, an intuitive generative mechanism is that all rows and columns are formed by a latent low-rank structure. There are at least two parallel but seemingly equivalent streams of works that are learning the low-rank formation of the matrix-variate data. The first is via matrix (or tensor) regularization with low-rank penalization, e.g., rank constraint or nuclear norm penalty, with or without noise. These results in great success in wide applications like matrix completion, compressed sensing and image localization (\cite{mao2019matrix} and \cite{mao2021Matrix}). The second line of work is the factor modeling which thinks all entries of a matrix observation are driven by a low-dimensional factor matrix, and the magnitude of each entry relies on the additive and/or interactive effect of the low-dimensional row features and column characteristics. In the present paper, we follow the factor modeling manner but a regularized optimization technique is used to adapt to the nonlinearity of our proposed model.

For ease of presentation, let $\{X_t = ( X_{ijt})_{p_1\times p_2}; 1\leq t\leq T\}$ be a matrix sequence (e.g., a matrix time series). There are also two ways to do matrix (or even tensor) factor analysis of $X_t$. The first manner is to ``flatten" each $X_t$ into a lengthy vector by stacking its columns or rows: the first step is learning the vector factor space with existing vector factor analysis methods (\cite{Bai2002De}, \cite{bai2012statistical}, \cite{fan2013large}, \cite{kong2017number}, \cite{pelger2019large}, \cite{Trapani2008Aran}, \cite{barigozzi2022testing}, \cite{Barigozzi2022StatisticalIF}), and the second step is recovering the row and column factor spaces from the vectorized factor space with certain restriction, e.g. the Kronecker structure, c.f, \cite{chen2024factor} and \cite{chen2021auto}. The second collection of papers  (\cite{wang2019factor}, \cite{WANG2022180}, \cite{chang2023modelling},  \cite{yu2022projected}, \cite{yuan2023two}, \cite{he2024matrixk}) directly model each $X_t$ with carefully designed additive and/or interactive row and column effects. It is worth noticing that \cite{wang2019factor} introduced the first matrix factor in the literature; \cite{yu2022projected} provided a projection approach to estimate the twisted row and column factor spaces which has, so far to the best of our knowledge, the fastest convergence rates in the class of PCA procedures; \cite{yuan2023two} presented the first two-way additive matrix factor model and \cite{zhang2024mod} combined the two-way additive and interactive components in generating the matrix entries recently; the seminal work of \cite{chen2022factor} extends to the general tensor data.

However, the above works are limited to single-type continuous variables. In practical applications, such as social science and biology, high-dimensional data sets often involve mixed-type variables. For instance, in the context of corporate operational performance analysis, there are not only continuous variables, such as return on equity and fixed asset turnover, but also categorical variables like industry type, enterprise nature (e.g., state-owned or private enterprise), and regional classification. These categorical indicators are often used to depict the fundamental characteristics and background information of a company. There are also some binary indicators to measure each corporate's social responsibility, such as whether the protection of employees' rights and interests is disclosed, and whether environmental sustainable development is disclosed. Collectively, the discrete variables encompassed in a company's economic indicators span multiple facets and play a pivotal role in portraying the company's economic standing, analyzing market trends, and devising business strategies. In networks analysis (\cite{Jing2021Community}, \cite{Jing2022Community}, \cite{chang2023modelling} and \cite{zhou2024attribute}) like international trading, the weight on each edge could be the trading direction (binary variables indicating import or export), the number of traded products (e.g. shoes, clothes, food) that are measured either in counts (count variables) or kilograms (continuous variables). For non-continuous variables, nonlinear models are usually employed for modeling, where closed-form expressions for estimating factors and their loadings do not exist, in contrast to the explicit principal component solution with or without iteration to the linear two-way factor analysis, c.f., \cite{Chen2023Statistical}, \cite{yu2022projected} and \cite{he2024matrixk}. Moreover, with mixed-type variables, the nonlinear structure varies depending on the type of variables, posing challenges for statistical inference and computation. To the best of our knowledge, there is still a lack of study on a general nonlinear model with factor structure for matrix sequences with mixed-type variables. This is a motivation and will be the first contribution of the present paper.

The literature recently starts to pay attention to the latent factor structure with single or mixed type variables in {\sc vector} factor modeling. \cite{CHENnonlinear} derived the asymptotic theory of the regression coefficients and the average partial effect for a class of generalized linear vector factor models that include logit, probit, ordered probit and Poisson models as special cases. Permitting general nonlinearity, \cite{WANG2022180} investigated the maximum likelihood estimation for the factors and loadings. \cite{Liu2023Gene} proposed a canonical exponential family model with factor structure for mixed-type variables. However, the likelihood approach for these nonlinear vector factor models can not be trivially extended to the matrix sequences since the Hessian matrix jointly in the row and column loadings and the factors are more complex. Extra manipulation is needed to feasibly derive the convergence property of the parameters related to the factor structure, especially for the asymptotic normality of the estimates, see Section \ref{method} for more details. This is a motivation and will be a second contribution to the present paper.

In this paper, we assume that each entry is generated by the following nonlinear generalized matrix factor model (GMFM):
\begin{equation}\label{GMFM}
    x_{ijt} \sim g_{ijt}\left(\cdot| \pi_{ijt}\right), i=1,\ldots,p_1,\,j=1,\ldots,p_2,\,t=1,\ldots,T.
\end{equation}
where $x_{ijt}$ is an observed entry lying in the $i$th row and $j$th column in sample $t$; $g_{ijt}(\cdot|\cdot)$ is some known probability (density or mass) function of $x_{ijt}$
allowed to vary across $i$, $j$ and $t$, permitting a mixture of distributions for each matrix; $\pi_{ijt} = r_i^{\prime}F_tc_j$ with $r_i$
being a $k_1$ dimensional vector of row loadings, $c_j$ being
a $k_2$ dimensional vector of column loadings, and $F_t$ being a $k_1\times k_2$ dimensional matrix of factors. Both
factors and loadings are unobservable and $k_1$ and $k_2$ are the numbers of row and column factors, respectively. In the currently studied linear matrix factor model, $\pi_{ijt}=r_i^{\prime}F_tc_j$ represents the conditional mean of $x_{ijt}$, while for non-continuous variables in the present paper, it might stand for the conditional log odds or log probabilities. Even for continuous variables, $\pi_{ijt}$ in model (\ref{GMFM}) can be a function of the conditional smoothly-winsorized-mean in the robust matrix factor model, c.f., He et al. (2024), which maps $g_{ijt}(\cdot | \cdot)$'s to the Huber loss function. Thus model (\ref{GMFM}) is a quite flexible nonlinear matrix factor model. Certainly, to identify $\left\{r_i,F_t,c_j\right\}$, one needs the identification constraints to form an identifiably feasible solution set.

In this paper, we consider the maximum (quasi-)likelihood estimation with rotational constraints for the factors and loadings of GMFM. We introduce a novel penalized likelihood function that not only covers the log-likelihood function and the identifiability restriction, but also includes an additional augmented Lagrangian term, which is carefully tuned to guarantee the local convexity of the negative Hessian matrix around the true parameters and hence a concave quasi-likelihood function with penalty in a neighborhood of the true parameters. This leads to a convenient derivation of the central limit theorems, and it is where the second contribution of the present paper comes from. This paper also establishes the convergence rates of the estimated factors and loadings. Our theory demonstrates that our estimators converge at rate $O_p\left(1/\min\left\{\sqrt{p_1p_2},\sqrt{p_1T},\sqrt{p_2T}\right\}\right)$
in terms of the averaged Frobenius norm. This rate surpasses $O_p(1/\min\{\sqrt{p_1p_2},\sqrt{T}\})$, the rate for generalized {\sc vector} factor analysis achieved through vectorizing ${X}_t$ by \cite{Liu2023Gene} and \cite{WANG2022180}, particularly when the sequence length $T$ is relatively short. It is also no slower than the rate, $O_p(\{1/\sqrt{p_1}+1/\sqrt{Tp_2}\}\vee \{1/\sqrt{p_2}+1/\sqrt{Tp_1}\})$, of the non-likelihood method $\alpha$-PCA in the seminal paper by \cite{Chen2023Statistical}, especially when the data matrix is not balanced. Furthermore, to consistently estimate the pair of numbers of the row and column factors, we present an information-type criterion under GMFM. This set of new results forms the third contribution of the present paper. Motivated by the challenges posed by nonlinear structures and mixed-type variables, we also present extended versions of the two-stage alternating maximization algorithm inspired by \cite{Liu2023Gene} and the minorization maximization algorithm inspired by \cite{CHENnonlinear}.

The rest of the paper is organized as follows. Section \ref{method} introduces the set up and the estimation methodology. Section \ref{result} presents the asymptotic theorems, and subsection \ref{computation} provides two computational algorithms. Section \ref{simulation} is devoted to extensive simulation studies. Section \ref{real data} analyzes high-tech company's operating performance. Section \ref{CD} concludes. All proofs are relegated to the supplementary materials.

\section{Methodology and Set Up} \label{method}

\subsection{Methodology}
With model (\ref{GMFM}), the (quasi) log-likelihood function is
\begin{align}
    L(X|r,f,c) = \sum_{i=1}^{p_1} \sum_{j=1}^{p_2} \sum_{t=1}^{T}l_{ijt}(r_i'F_tc_j), \label{Model:Our}
\end{align}
where $l_{ijt}(\pi_{ijt}) = \log g_{ijt}\left(x_{ijt}|\pi_{ijt}\right)$ with $\pi_{ijt} = r_i'F_tc_j$; $r = \left(r_1',\ldots,r_{p_1}'\right)'$ is a $p_1 k_1$ vector, $c = \left(c_1',\ldots,c_{p_2}'\right)'$ is a $p_2 k_2$ vector, and $f = \left(f_1',\ldots,f_T'\right)'$ with $f_t = vec\{F_t\}$ is a $T k_1k_2$ vector. The entries of $X$ can be continuous variables, count variables, binary variables, and so on. Representative examples are as follows.

\noindent \textbf{Example 1 (Linear)} $l_{ijt}(\pi_{ijt}) = -\frac{1}{2}\left(x_{ijt} - r_i'F_tc_j\right)^2$ is a likelihood function of Gaussian random variables and a quasi-likelihood function for other continuous random variables with homoscedasticity.

\noindent\textbf{Example 2 (Poisson)} $l_{ijt}(\pi_{ijt}) = -e^{r_i'F_tc_j}+kr_i'F_tc_j-\log k!$ as $P\left(X_{ijt} = k\right) = e^{-\lambda}\lambda^k/k!$ and $\lambda = e^{r_i'F_tc_j}$.

\noindent\textbf{Example 3 (Probit)} $l_{ijt}(\pi_{ijt}) = x_{ijt}\log \Phi(r'_iF_tc_j) + (1-x_{ijt})\log\left(1-\Phi(r'_iF_tc_j)\right)$, where $\Phi(\cdot)$ is the cumulative distribution function of a standard normal random variable.

\noindent\textbf{Example 4 (Logit)} $l_{ijt}(\pi_{ijt}) = x_{ijt}\log \Psi(r'_iF_tc_j) + (1-x_{ijt})\log\left(1-\Psi(r'_iF_tc_j)\right)$, where $\Psi(\cdot)$ is the cumulative distribution function of the logistic distribution.

\noindent\textbf{Example 5 (Tobit)} Let $x_{ijt} = x^*_{ijt} I(x^*_{ijt}>0)$ where $x_{ijt}^* = r_i'F_tc_j+e_{ijt}$ and $e_{ijt}$ is $N(0,1)$.
$l_{ijt}(\pi_{ijt}) = -\frac{1}{2}(x_{ijt}-r'_iF_tc_j)^2 I(x_{ijt}>0) + \log\left(1-\Phi(r'_iF_tc_j)\right) I(x_{ijt}=0)$, where $ I(\cdot) $ is the indicator function.

Let ${R} = \left(r_1,\ldots,r_{p_1}\right)'$ be the $p_1\times k_1$ row factor loading matrix, ${C} =\left(c_1,\ldots,c_{p_2}\right)'$ be the $p_2\times k_2$  column factor loading matrix, ${F}_t$ be a $k_1\times k_2$ common factor matrix. Denote the parameters of interest by the vector $\theta = \left(r',f',c'\right)'$ and their true values by
${r}_i^0$, ${F}_t^0$ and ${c}_j^0$, respectively. To well define a high-dimensional parameter space $\Theta$ of $\theta$, except the boundedness condition for each row of $R$ and $C$ and the factor matrix $F_t$ in Assumption \ref{ass1}, one needs constraints to identify the parameters.
Note that for any $k_1\times k_1$, $k_2\times k_2$ invertible
matrix $G_1$ and $G_2$, ${R}G_1$, $G_1^{-1}{F}_tG_2^{-1}$ and ${C}G^{\prime}_2$ give the same likelihood as ${R}$, ${F}_t$ and ${C}$ do. To uniquely determine ${R}$, ${F}_t$, and ${C}$, without loss of
generality, we impose the identifiability restriction as follows.
\begin{eqnarray}\label{constraint}
\left\{\begin{array}{lll} (3.1) \ \ \frac{{R}'{R}}{p_1} = \mathbb{I}_{k_1}, \frac{{C}'{C}}{p_2} = \mathbb{I}_{k_2}, \\
(3.2) \ \ \sum_{t=1}^T\frac{{F}_t{F}'_t}{T} \ \mbox{and} \ \sum_{t=1}^T\frac{{F}'_t{F}_t}{T} \ \mbox {diagonal matrices}, \\
(3.3) \ \ \mbox{the first nonzero element in each column of} \ R \ \mbox{and} \ C \ \mbox{is positive}.
\end{array}\right.
\end{eqnarray}

The conditions (3.1) and (3.2) are frequently used in the literature of matrix factor models, c.f., \cite{he2024matrixk}. Without condition (3.3), results in factors and loadings can only be unambiguously determined up to a sign matrix (diagonal matrix with $1$ and $-1$ on the diagonals) transformation. Then a naturally identified estimate of the vector of factor and loading parameters is the maximizer of (\ref{Model:Our}) subject to the constraints in (\ref{constraint}). It can be easily shown this constrained maximizer is equivalent to the solution to the following augmented Lagrange function up to the sign undeterminance:
\begin{eqnarray}\label{EQ:ConLM}
   \max_{r, f, c\in \Theta} Q(r,f,c) &=& L(X |r,f,c) + P(r,f,c), \\
   P(r, f, c) &=& P_1(r, f, c)+P_2(r, f, c)+P_3(r, f, c),
\end{eqnarray}
where
 \begin{eqnarray}
      P_1({\theta}) &=& -b_1p_1p_2T\Bigg[\frac{1}{2p_1^2}\sum_{p=1}^{k_1}\sum_{q>p}^{k_1}\left(\sum_{i=1}^{p_1}r_{ip}r_{iq}\right)^2 + \frac{1}{8p_1^2}\sum_{k=1}^{k_1}\left(\sum_{i=1}^{p_1}r_{ik}^2-p_1\right)^2  \nonumber\\
      &&+ \frac{1}{2T^2}\sum_{p=1}^{k_1}\sum_{q>p}^{k_1}\left(\sum_{t=1}^T{F}_{tp\cdot}{F}_{tq\cdot}^{\prime}\right)^2\Bigg],\\
      P_2({\theta}) &=& -b_2p_1p_2T\Bigg[\frac{1}{2p_2^2}\sum_{p=1}^{k_2}\sum_{q>p}^{k_2}\left(\sum_{j=1}^{p_1}c_{jp}c_{jq}\right)^2 + \frac{1}{8p_2^2}\sum_{k=1}^{k_2}\left(\sum_{j=1}^{p_2}c_{jk}^2-p_2\right)^2  \nonumber \\
      &&+ \frac{1}{2T^2}\sum_{p=1}^{k_2}\sum_{q>p}^{k_2}\left(\sum_{t=1}^T{F}_{t\cdot p}^{\prime}{F}_{t\cdot q}\right)^2\Bigg],\\
  P_3({\theta}) &=& -b_3p_2\sum_{t=1}^T\Bigg[\frac{1}{2p_1}\sum_{p=1}^{k_1}\sum_{q=1}^{k_2}\left(\sum_{i=1}^{p_1}\left(\frac{r_{ip}^2-1}{2}f_{t,pq} + \sum_{k\neq p}^{k_1}r_{ip}r_{ik}f_{t,kq}\right)\right)^2\Bigg].
\end{eqnarray}
where $b_1$, $b_2$ and $b_3$ are positive Lagrange multipliers. The penalty terms $P_1(\theta)$ and $P_2(\theta)$ are associated with the conditions (3.1) and (3.2). The additional augmented term $P_3(\theta)$ is constructed to ensure a locally positive definite property of the negative Hessian matrix of the penalized likelihood function $Q(\theta)$ around the true parameter vector $\theta^0$, by cleverly trading off the non-diagonal blocks of the scaled expectation of the Hessian matrix of $L(\theta)$. This property shows a desirable geometric concavity of the augmented Lagrange function $Q(\cdot)$ and leads to an easy way to derive the second-order asymptotics, in particular the central limit theorems of the estimated parameters. Without this newly added term, simply introducing \(P_1(\theta) + P_2(\theta)\), might not result in a desirable landscape around the local minima of \(Q(\theta)\). In the sequel, we shall consider the estimated factors and loadings as the solution to maximizing $Q(r,f,c)$ and obtain the asymptotic results for matrix factor analysis. The equivalence between (\ref{Model:Our})-(\ref{constraint}) and (\ref{EQ:ConLM}) is due to the non-positiveness of $P(\theta)$ and the fact that $Q(\theta)$ is maximized if and only if $P(\theta)=0$.

Throughout the paper, $(p_1,p_2,T) \to\infty$ means $p_1,p_2$ and $T$ going to infinity jointly,  ``w.p.a.1" is a short abbreviation of ``with probability approaching $1$". For vectors, $\Vert \cdot\Vert$ denotes the Euclidean norm. For matrix $A$, $\rho_{\min}(A)$ is its smallest eigenvalue,
and $\Vert A\Vert$, $\Vert A\Vert_F$, $\Vert A\Vert_1$, $\Vert A\Vert_{\infty}$ and $\Vert A\Vert_{\max}$ denotes the spectral norm, Frobenius norm, $1$-norm, infinity norm and max norm,
respectively. When $A$ has $Tk$ rows, divide $A$ into $T$ blocks with each block containing $k$ rows, and let $[A]_{tq}$ denote the $q$th
row in the $t$th block and $[A]_t = \left([A]'_{t1},\ldots,[A]'_{tk}\right)$ is the $t$th block. Define $L_{p_1p_2T} = \min\{\sqrt{p_1p_2},\sqrt{p_1T},\sqrt{p_2T}\}$.

\subsection{Set Up Assumptions}\label{setup}

In this section, we give some technical assumptions that are used to establish the asymptotic theories. In the sequel, $M$ will be a generic constant that might vary from line to line.

\begin{assumption}\label{ass1}
\begin{enumerate}
\item $\left\Vert r_i^0\right\Vert\leq M$, $\left\Vert c_j^0\right\Vert\leq M$ and $\left\Vert {F}_t^0\right\Vert\leq M$ for all $i, j$ and $t$.
\item $\frac{1}{T}\sum_{t=1}^T{F}^0_t{F}_t^{0\prime} = \diag\left(\sigma_{T1},\ldots,\sigma_{Tk_1}\right)$ with $\sigma_{T1}\geq\cdots\geq\sigma_{Tk_1}$, and $\sigma_{Tk}\to\sigma_{k}$ as $T\to\infty$ for $k=1,\ldots,k_1$ with $\sigma_1>\cdots>\sigma_{k_1}>0$.
$\frac{1}{T}\sum_{t=1}^T{F}^{0\prime} _t{F}_t^{0} = \diag\left(\sigma'_{T1},\ldots,\sigma'_{Tk_2}\right)$ with $\sigma'_{T1}\geq\cdots\geq\sigma'_{Tk_2}$, and $\sigma'_{Tl}\to\sigma_{l}$ as $T\to\infty$ for $l=1,\ldots,k_2$ with $\sigma'_1>\cdots>\sigma'_{k_2}>0$.
\end{enumerate}
\end{assumption}

Assumption 1-1 indicates that ${r}_i^0, {F}_t^0, {c}_j^0$
 are uniformly bounded.
As explained in \cite{WANG2022180},
the compactness of parameter space is a commonly encountered trait in nonlinear models, see more examples in \cite{Newey1986LargeSE}, \cite{Jennrich1969} and \cite{Wu1981}. Assumption 1-2 assumes asymptotic distinct eigenvalues for $\frac{1}{T}\sum^T_{t=1}F_t^0F_t^{0\prime}$ and $\frac{1}{T}\sum^T_{t=1}F_t^{0\prime}F_t^0$ so that the eigenvectors can be uniquely determined. This is the same as Assumption B in \cite{yu2022projected}.

Let $\partial_{\pi}l_{ijt}(\pi_{ijt})$, $\partial_{\pi^2}l_{ijt}(\pi_{ijt})$ and $\partial_{\pi^3}l_{ijt}(\pi_{ijt})$ be the first, second and third order derivative of $l_{ijt}(\theta)$ evaluated at $\pi_{ijt}$, respectively.
When these derivatives are evaluated at $\pi_{ijt}^0$, we suppress the argument and simply write them as $\partial_{\pi}l_{ijt}$, $\partial_{\pi^2}l_{ijt}$ and $\partial_{\pi^3}l_{ijt}$.

\begin{assumption}\label{ass2}
\begin{enumerate}
\item $l_{ijt}(\theta)$  is three times differentiable.
\item  There exists $b_U > b_L > 0$ such that $b_L \leq -\partial_{\pi^2}l_{ijt}(\pi_{ijt}) \leq b_U$.
\item $\left\vert \partial_{\pi^3}l_{ijt}(\pi_{ijt})\right\vert\leq b_U$ within a compact space of $\pi_{ijt}$.
\end{enumerate}
\end{assumption}

Assumption
2-1 imposes a smoothness condition on the log-likelihood function. Assumptions 2-2 and 2-3 assume
that the second-order derivative of the log-likelihood function is bounded from below and above, and the third-order derivative is bounded away from infinity. The boundedness of the second and the third-order derivative is used to control
the remainder term in the expansion of the first-order condition. The boundedness from below of the second order derivative together with the boundedness of $\pi_{ijt}$ are used to show the consistency of the estimated factors and loadings. It can be easily verified that commonly used models, such as Linear, Logit, Probit, Poisson, and Tobit, satisfy Assumption 2.

\begin{assumption}\label{ass3}
\begin{enumerate}
\item $\E\left(\left\vert \partial_{\pi}l_{ijt}\right\vert^{\xi}\right)\leq M$ for some $\xi>14$ and all $i,j$ and $t$.

\item For any $i,j,t$, $\sum_{s=1}^T\sum_{l=1}^{p_1}\sum_{k=1}^{p_2}\left\vert \E\partial_{\pi}l_{ijt}\partial_{\pi}l_{lks} \right\vert \leq M$.

\item For every $(t,s)$,
$\E\left[p_1^{-1/2}p_2^{-1/2}\sum_{i=1}^{p_1}\sum_{j=1}^{p_2}\sum_{k=1}^{p_2}\left(\partial_{\pi}l_{ijt}\partial_{\pi}l_{iks}-\E\partial_{\pi}l_{ijt}\partial_{\pi}l_{iks}\right)\right]^2\leq M$,\\
$\E\left[p_1^{-1/2}p_2^{-1/2}\sum_{i=1}^{p_1}\sum_{l=1}^{p_1}\sum_{j=1}^{p_2}\left(\partial_{\pi}l_{ijt}\partial_{\pi}l_{ljs}-\E\partial_{\pi}l_{ijt}\partial_{\pi}l_{ljs}\right)\right]^2\leq M$,\\
For every $(j,k)$,
$\E\left[p_1^{-1/2}T^{-1/2}\sum_{i=1}^{p_1}\sum_{t=1}^{T}\sum_{s=1}^{T}\left(\partial_{\pi}l_{ijt}\partial_{\pi}l_{iks}-\E\partial_{\pi}l_{ijt}\partial_{\pi}l_{iks}\right)\right]^2\leq M$.
\end{enumerate}
\end{assumption}

Notice that for linear model, $\partial_{\pi}l_{ijt}$ is the error term and $\partial_{\pi^2}l_{ijt}$ is a constant. Many papers require $\partial_{\pi}l_{ijt}$ to be conditionally independent for using the Hoeffding's inequality in the proofs, such as \cite{he2023Iterative}. This paper considers a more general situation: the distributions of $x_{ijt}$ are allowed to be heterogeneous over $i$,$j$ and $t$, and to have limited cross-row (-column) and cross-sample (e.g, serial) dependence of $x_{ijt}$ conditionally.
Assumption \ref{ass3} gives some moment conditions for $\{\partial_{\pi}l_{ijt}\partial_{\pi}l_{ijs}; 1\leq i\leq p_1, 1\leq j\leq p_2, 1\leq t\leq T\}$. They are satisfied when, for example, $\left\{\partial_{\pi}l_{ijt}\right\}$ are cross-sectionally and temporally weakly dependent conditional on $\theta\in\Theta$.

\begin{assumption}\label{ass4}
For some $\zeta>2$,
\begin{eqnarray*}
\E\left(p_1^{-1}\sum_{i=1}^{p_1}\left\Vert \frac{1}{\sqrt{p_2T}}\sum_{j=1}^{p_2}\sum_{t=1}^T\partial_{\pi}l_{ijt}F_t^0c^0_j\right\Vert^{\zeta}\right)&\leq & M, \\
\E\left(p_2^{-1}\sum_{j=1}^{p_2}\left\Vert \frac{1}{\sqrt{p_1T}}\sum_{i=1}^{p_1}\sum_{t=1}^T\partial_{\pi}l_{ijt}F_t^{0\prime}r^0_i\right\Vert^{\zeta}\right)&\leq & M, \\
\E\left(T^{-1}\sum_{t=1}^{T}\left\Vert \frac{1}{\sqrt{p_1p_2}}\sum_{i=1}^{p_1}\sum_{j=1}^{p_2}\partial_{\pi}l_{ijt}c^0_j\otimes r^0_i\right\Vert^{\zeta}\right)&\leq & M.
\end{eqnarray*}
\end{assumption}

Assumption \ref{ass4} is for the first derivatives of the log-likelihood function. Again, it is satisfied when $\left\{\partial_{\pi}l_{ijt}\right\}$ are cross-sectionally and temporally weakly dependent conditional on $\theta\in\Theta$. To present the next assumption on the moments of the second derivatives of the log-likelihood function, we introduce some more notations. Define
\begin{align*}
        \mathcal{H}_{Lrr'} &=\left[\sum_{j=1}^{p_2}\sum_{t=1}^T\partial_{\pi^2}l_{ijt}F_t^0c_j^0c_j^{0\prime}F_t^{0\prime}\right]_{i=1}^{p_1}, \
    \mathcal{H}_{Lff'} = \left[\sum_{i=1}^{p_1}\sum_{j=1}^{p_2}\partial_{\pi^2}l_{ijt}c_j^0\otimes r_i^0c_j^{0\prime}\otimes r_i^{0\prime}\right]_{t=1}^{T},\\
    \mathcal{H}_{Lcc'} &=\left[\sum_{i=1}^{p_1}\sum_{t=1}^T\partial_{\pi^2}l_{ijt}F_t^{0\prime}r_i^0r_i^{0\prime}F_t^{0}\right]_{j=1}^{p_2}.
  \end{align*}

\begin{assumption}\label{ass5}
\begin{enumerate}
\item Recall the notation $[A]_l$ for some matrix $A$,
\begin{eqnarray*}    \E\Big\Vert\frac{1}{\sqrt{p_1p_2T}}\sum_{i=1}^{p_1}\sum_{j=1}^{p_2}\sum_{t=1}^T\left(\frac{1}{p_2T}[\mathcal{H}_{Lrr'}]_i\right)^{-1}\partial_{\pi}l_{ijt}F_t^0c_j^0r_i^{0\prime}\Big\Vert^2 &\leq & M,\\
\E\Big\Vert\frac{1}{\sqrt{p_1p_2T}}\sum_{i=1}^{p_1}\sum_{j=1}^{p_2}\sum_{t=1}^T\left(\frac{1}{p_1p_2}[\mathcal{H}_{Lff'}]_t\right)^{-1}\partial_{\pi}l_{ijt}c_j^0\otimes r_i^{0}f_t^{0\prime}\Big\Vert^2 & \leq & M,\\
\E\Big\Vert\frac{1}{\sqrt{p_1p_2T}}\sum_{i=1}^{p_1}\sum_{j=1}^{p_2}\sum_{t=1}^T\left(\frac{1}{p_1T}[\mathcal{H}_{Lcc'}]_j\right)^{-1}\partial_{\pi}l_{ijt}F_t^{0\prime}r_i^0c_j^{0\prime}\Big\Vert^2 &\leq & M.
\end{eqnarray*}

\item  For any $k$, $s$ and $l$,

$\E\Big\Vert\frac{1}{\sqrt{p_1p_2T}}\sum_{i,j,t}\left(\frac{1}{T}\sum_{t=1}^{T}\left(\partial_{\pi^2}l_{ikt}F_t^{0\prime} r_i^0c_k^{0\prime}F_t^{0\prime} + \partial_{\pi}l_{ikt}F_{t}^{\prime}\right)\right)\left(\frac{1}{p_2T}[\mathcal{H}_{Lrr'}]_i\right)^{-1}\\\times\left(\partial_{\pi}l_{ijt}F_t^0c_j^0\right)\Big\Vert^2\leq M$,

$\E\Big\Vert\frac{1}{\sqrt{p_1p_2T}}\sum_{i,j,t}\left(\frac{1}{p_2}\sum_{j=1}^{p_2}\left(\partial_{\pi^2}l_{ijs}c_j^0\otimes r_i^0c_j^{0\prime}F_s^{0\prime} + \partial_{\pi}l_{ijs}c_j^0\otimes \mathbb{I}_{k_1}\right)\right)\left(\frac{1}{p_2T}[\mathcal{H}_{Lrr'}]_i\right)^{-1}\\
\times \Big(\partial_{\pi}l_{ijt}F_t^0c_j^0\Big)\Big\Vert^2\leq M$,

$\E\Big\Vert\frac{1}{\sqrt{p_1p_2T}}\sum_{i,j,t}\left(\frac{1}{p_1}\sum_{j=1}^{p_2}\left(\partial_{\pi^2}l_{ljt}F_t^{0}c_j^0c_j^{0\prime}\otimes r_l^{0\prime} + \partial_{\pi}l_{ljt}c_j^{0\prime}\otimes \mathbb{I}_{k_1}\right)\right)\left(\frac{1}{p_1p_2}[\mathcal{H}_{Lff'}]_t\right)^{-1}\\
\times\left(\partial_{\pi}l_{ijt}c_j^0\otimes r_i^0\right)\Big\Vert^2\leq M$,

$\E\Big\Vert\frac{1}{\sqrt{p_1p_2T}}\sum_{i,j,t}\left[\frac{1}{p_2}\sum_{j=1}^{p_2}\left(\partial_{\pi^2}l_{ikt}F_t^{0\prime}r_i^0c_k^{0\prime}\otimes r_i^{0\prime} + \partial_{\pi}l_{ikt}\mathbb{I}_{k_2}\otimes r_i^{0\prime}\right)\right]\left(\frac{1}{p_1p_2}[\mathcal{H}_{Lff'}]_t\right)^{-1}\\
\times\left(\partial_{\pi}l_{ijt}c_j^0\otimes r_i^0\right)\Big\Vert^2\leq M$,

$\E\Big\Vert\frac{1}{\sqrt{p_1p_2T}}\sum_{i,j,t}\left[\frac{1}{p_1}\sum_{i=1}^{p_1}\left(\partial_{\pi^2}l_{ijs}{c}^0_j\otimes{r}^0_i{r}^{0\prime}_i{F}^0_s+ \partial_{\pi}l_{ijs}\mathbb{I}_{ k_2}\otimes{r}^0_i\right)\right]\left(\frac{1}{p_1T}[\mathcal{H}_{Lcc'}]_j\right)^{-1}\\
\times\left(\partial_{\pi}l_{ijt}F_t^{0\prime}r_i^0\right)\Big\Vert^2\leq M$,

$\E\Big\Vert\frac{1}{\sqrt{p_1p_2T}}\sum_{i,j,t}\left[\frac{1}{T}\sum_{t=1}^{T}\left(\partial_{\pi^2}l_{ljt}{F}^0_t{c}^0_j{r}^{0\prime}_l{F}^0_t+ \partial_{\pi}l_{ljt}{F}^0_t\right)\right]\left(\frac{1}{p_1T}[\mathcal{H}_{Lcc'}]_j\right)^{-1}\\
\times\left(\partial_{\pi}l_{ijt}F_t^{0\prime}r_i^0\right)\Big\Vert^2\leq M$.

\item For any $i$ and some positive definite matrices $\Sigma_{iR}$ and $\Omega_{iR}$,
\begin{eqnarray*}
    -(p_2T)^{-1}\sum_{j=1}^{p_2}\sum_{t=1}^T\partial_{\pi^2}l_{ijt}F_t^0c_j^0c_j^{0\prime}F_t^{0\prime}&\xrightarrow{P}& \Sigma_{iR},\\
    (p_2T)^{-1/2}\sum_{j=1}^{p_2}\sum_{t=1}^T\partial_{\pi}l_{ijt}F_t^0c_j^0&\xrightarrow{d}& \mathcal{N}(0,\Omega_{iR}).
\end{eqnarray*}

For any $j$ and some positive definite matrices $\Sigma_{jC}$ and $\Omega_{jC}$,
     \begin{eqnarray*}
         -(p_1T)^{-1}\sum_{i=1}^{p_1}\sum_{t=1}^T\partial_{\pi^2}l_{ijt}F_t^{0\prime}r_i^0r_i^{0\prime}F_t^{0}&\xrightarrow{P}& \Sigma_{jC}, \\ (p_1T)^{-1/2}\sum_{i=1}^{p_1}\sum_{t=1}^T\partial_{\pi}l_{ijt}F_t^{0\prime}r_i^0&\xrightarrow{d} & \mathcal{N}(0,\Omega_{jC}).
     \end{eqnarray*}

 For any $t$ and some positive definite matrices $\Sigma_{tF}$ and $\Omega_{tF}$,
     \begin{eqnarray*}
        -(p_1p_2)^{-1}\sum_{i=1}^{p_1}\sum_{j=1}^{p_2}\partial_{\pi^2}l_{ijt}c_j^{0}\otimes r_i^0r_i^{0\prime}\otimes c_j^{0\prime}&\xrightarrow{P}& \Sigma_{tF},\\
        (p_1p_2)^{-1/2}\sum_{i=1}^{p_1}\sum_{j=1}^{p_2}\partial_{\pi}l_{ijt}c_j^{0}\otimes r_i^0&\xrightarrow{d}& \mathcal{N}(0,\Omega_{tF}).
     \end{eqnarray*}

\end{enumerate}
\end{assumption}

Though cumbersome, it can be easily verified that Assumption \ref{ass5} is also satisfied when $\left\{\partial_{\pi}l_{ijt}\right\}$ are cross-sectionally and temporally
weakly dependent conditional on $\theta \in \Theta$.

\begin{assumption}\label{ass6} $\frac{(p_1p_2T)^{3/\xi}(p_1+p_2+T)^{1/\zeta}}{L_{p_1p_2T}}\to 0$, as $(p_1,p_2,T)\to \infty$.
\end{assumption}

Assumption 6 is sort of a balance condition for $p_1$, $p_2$ and $T$, meaning that each dimension size can not completely dominate the other two, which is controlled by $\xi$ and $\zeta$. It is
weaker when $\xi$ and $\zeta$ are larger. For example, when $\partial_{\pi}l_{ijt}$ has subgaussian tails, $\xi$ can be arbitrarily large. And when $\partial_{\pi}l_{ijt}$'s are conditionally Gaussian as a typical case in the linear model, $\zeta$ can be arbitrarily large.

\section{Asymptotic Theory} \label{result}

Let $B(\mathcal{D})=\{(r, f, c): \Vert {r}\Vert_{\infty}+\Vert f\Vert_{\infty} +\Vert {c}\Vert_{\infty} \leq \mathcal{D}\}$ for some $\mathcal{D}$ large enough, such that the true
parameter ${\theta}^0$
lies in the interior of $B(\mathcal{D})$. Before stating the theoretical results, we give some more notations. Define $\widehat{r} = (\widehat{r}_1',\ldots,\widehat{r}_{p_1}')'$, $\widehat{c} = (\widehat{c}_1',\ldots,\widehat{c}_{p_2}')'$, and  $\widehat{f} = (\widehat{f}_1',\ldots,\widehat{f}_T')'$ as the solution of maximizing
$Q$ within $B(\mathcal{D})$, where $\widehat{f}_t = \text{vec}\{\widehat{F}_t\}$. Let $\widehat{\pi}_{ijt} = \widehat{r}_i\widehat{F}_t\widehat{c}_j$, $\widehat{{\theta}} = (\widehat{r}',\widehat{f}',\widehat{c}')'$, $\widehat{R} = (\widehat{r}_1,\ldots,\widehat{r}_{p_1})'$, $\widehat{C} = (\widehat{c}_1,\ldots,\widehat{c}_{p_2})'$. Let $S({\theta}) = \partial_{{\theta}}Q({\theta})$, $S_{r}({\theta}) = \partial_{r}Q({\theta})$, $S_{c}({\theta}) = \partial_{c}Q({\theta})$ and $S_{f}({\theta}) = \partial_{f}Q({\theta})$ be the score functions. Let $\mathcal{H}(\theta) = \partial_{\theta\theta'}Q(\theta)$ be the Hessian matrix. The decomposition of $\mathcal{H}(\theta)$ and the expression of each
component is presented in Appendix A in the supplementary material. We suppress the argument when $S(\theta)$ and $\mathcal{H}(\theta)$ are evaluated at $\theta^0$, i.e. $S = S(\theta^0)$ and $\mathcal{H} = \mathcal{H}(\theta^0)$.

\subsection{Convergence Rates}

We first provide a consistent result for the estimates of factors and loadings in terms of their average Euclidean norm (or equivalently the Frobenious norm of the estimated row and column factor loading matrices and the factor matrix).

\begin{proposition}
    (Average Consistency). Under Assumptions \ref{ass1}-\ref{ass3}, as $(p_1,p_2,T)\to\infty$,
      $$
      \frac{1}{p_1}\Vert \widehat{r}-r^0\Vert^2 + \frac{1}{p_2}\Vert\widehat{c} - c^0\Vert^2 + \frac{1}{T}\Vert\widehat{f} - f^0\Vert^2 = O_p\left(L_{p_1p_2T}^{-1}\right).
      $$
    \end{proposition}

To derive finer convergence rates and the limit distributions of the estimated parameters, we need to utilize the first-order condition $S(\widehat{\theta}) = 0$. The following proposition  demonstrates  that $S(\widehat{\theta}) = 0$ w.p.a.1.

\begin{proposition}
    Under Assumptions \ref{ass1}-\ref{ass4}, and Assumption \ref{ass6}, $S(\widehat{\theta}) = 0$ w.p.a.1.
\end{proposition}

Proposition 2 is nontrivial because the dimension of \(\widehat{\theta}\) increases with \(p_1\), \(p_2\) and \(T\). In a fixed-dimensional parameter space, the consistency of the estimated parameters, coupled with the assumption that the true parameters lie in the interior of the parameter space, ensures that the estimated parameters also lie in the interior. Consequently, the first-order conditions are satisfied. However, when the dimension of the parameter space increases with \(p_1\), \(p_2\), and \(T\), the average consistency of \(\widehat{\theta}\) as established in Proposition 1 is insufficient to ensure that \(\widehat{\theta}\) remains an interior point of the parameter space. In this context, the uniform consistency of \(\widehat{\theta}\) is required.

\begin{proposition}
   (Uniform Consistency). Under Assumptions \ref{ass1}-\ref{ass4} and \ref{ass6},
   $$
   \Vert \widehat{r}-r^0\Vert_{\infty} + \Vert \widehat{c}-c^0\Vert_{\infty} +  \Vert \widehat{f} - f^0\Vert_{\infty} = O_p\left(\frac{(p_1p_2T)^{3/\xi}}{L_{p_1p_2T}}(p_1+p_2+T)^{1/\zeta}\right).
   $$
\end{proposition}

Note that $\xi$ and $\zeta$ could be arbitrarily large in some examples as demonstrated below Assumption \ref{ass6}, and in such cases
$$
\Vert \widehat{r}-r^0\Vert_{\infty} + \Vert \widehat{c}-c^0\Vert_{\infty} +  \Vert \widehat{f} - f^0\Vert_{\infty} = O_p\left(L^{-1}_{p_1p_2T}\right).
$$

Now we utilize the first-order conditions. Using the integral form of the mean value theorem for vector-valued
functions to expand the first order conditions, we have $0 = S(\widehat{{\theta}}) = S+\widetilde{\mathcal{H}}\left(\widehat{{\theta}} - {\theta}^0\right)$, where $\widetilde{\mathcal{H}} = \int_0^1 \mathcal{H}\left({\theta}^0 + s\left(\widehat{{\theta}} - {\theta}^0\right)\right)ds$. It shows that
\begin{eqnarray}
    \left(\begin{array}{c}
         p_1^{-1/2}(\hat{r} - r^0)  \\
         T^{-1/2}(\hat{f} - f^0) \\
    p_2^{-1/2}(\hat{c} - c^0)\end{array}\right) = (p_1p_2T)^{-1/2}\left(-D_{p_2Tp_1}^{-1/2}\widetilde{\mathcal{H}}D_{p_2Tp_1}^{-1/2}\right)^{-1}D_{p_2Tp_1}^{-1/2}S,\label{EQ:thetahat-theta}
\end{eqnarray}
where
$$
D_{p_1Tp_2} = \diag\{p_1\mathbb{I}_{p_1k_1},T\mathbb{I}_{Tk_1k_2},p_2\mathbb{I}_{p_2k_2}\}, D_{p_2Tp_1} = \diag\{p_2T\mathbb{I}_{p_1k_1},p_1p_2\mathbb{I}_{Tk_1k_2},p_1T\mathbb{I}_{p_2k_2}\}.
$$
It is straightforward to see that $\| D_{p_2Tp_1}^{-1/2}S\| = O_p((p_1+p_2+T)^{1/2})$. Utilizing the local positive-definiteness of $-\mathcal{H}({\theta})$ aided by the carefully designed augmented Lagrangian terms, we show in the supplementary material (Lemma A.3) that the largest eigenvalue of $(-D_{p_2Tp_1}^{-1/2}{\mathcal{H}}(\theta)D_{p_2Tp_1}^{-1/2})^{-1}$ is $O_p(1)$ in the neighborhood $B(\mathcal{D})\cap \| D_{p_1Tp_2}^{-1/2}\left({\theta} - {\theta}^0\right)\|\leq m$ for some $m > 0$. Since $\widehat{\theta}$ lies in $B(\mathcal{D})\cap \| D_{p_1Tp_2}^{-1/2}\left({\theta} - {\theta}^0\right)\|\leq m$ w.p.a.1, this implies that $\|(-D_{p_2Tp_1}^{-1/2}\widetilde{\mathcal{H}}D_{p_2Tp_1}^{-1/2})^{-1}\|$ is $O_p(1)$. Then we have the following strengthened results.

\begin{theorem}\label{th1}
     (Average Rate). Under Assumptions \ref{ass1}-\ref{ass4} and Assumption \ref{ass6},
       $ \frac{1}{p_1}\Vert \widehat{r}-r^0\Vert^2 + \frac{1}{p_2}\Vert\widehat{c} - c^0\Vert^2 + \frac{1}{T}\Vert\widehat{f} - f^0\Vert^2 = O_p\left(L_{p_1p_2T}^{-2}\right)$.
    \end{theorem}

Comparable outcomes have been established for continuous variables. For example, the averaged convergence rate of ${\theta}$ by the least square estimate given in \cite{he2023Iterative} and the PE estimate given in \cite{yu2022projected}, both reached $L^{-1}_{p_1p_2T}$.
When the matrices are vectorized and the method for the generalized vector factor model is implemented as in \cite{Liu2023Gene} and \cite{WANG2022180}, the averaged convergence rate of estimating
the loading space spanned by ${C}\otimes{R}$ is $\min\{O_p(1/\sqrt{T},1/\sqrt{p_1p_2})\}$, which is no faster than ours, especially in scenarios where $p_1p_2$ outweighs $T$. It is also no slower than the rate, $O_p(\{1/\sqrt{p_1}+1/\sqrt{Tp_2}\}\vee \{1/\sqrt{p_2}+1/\sqrt{Tp_1}\})$, of the non-likelihood method $\alpha$-PCA in the seminal paper by \cite{Chen2023Statistical}, especially when the data matrix is not balanced. Yet extra effort has to be carried to separate $R$ and $C$ from their twisted Kronecker product.

\subsection{Limit distributions}\label{limitdis}
Now we proceed to the limit distributions of the estimated factors and loadings. Expanding the first-order conditions to a higher order, we have
$$0 = S(\widehat{{\theta}}) = S+\mathcal{H}(\widehat{{\theta}} - {\theta}^0) + \frac{1}{2}E,$$ where $E = (E_r',E_f',E_c')'$. $E_r$, $E_f$ and $E_c$ are $p_1k_1$, $Tk_1k_2$ and $p_2k_2$ dimensional with element $E_{r,ik} = (\hat{\theta}-\theta^0)'\partial_{\theta\theta'r_{ik}}Q(\theta_{ik}^*)(\hat{\theta}-\theta^0)$, $E_{f,tk} = (\hat{\theta}-\theta^0)'\partial_{\theta\theta'f_{tk}}Q(\theta_{tk}^*)(\hat{\theta}-\theta^0)$ and $E_{c,jk} = (\hat{\theta}-\theta^0)'\partial_{\theta\theta'c_{jk}}Q(\theta_{jk}^*)(\hat{\theta}-\theta^0)$,  respectively. $\theta_{ik}^*$, $\theta_{tk}^*$ and $\theta_{jk}^*$
are linear combinations of $\hat{\theta}$ and $\theta^0$. Thus
\begin{eqnarray*}
    \widehat{{\theta}} - {\theta}^0 &=& -\mathcal{H}^{-1}S - \frac{1}{2}\mathcal{H}^{-1}E, \\
    \hat{r}_i - r_i^0 &=& \left[\widehat{{\theta}} - {\theta}^0\right]_i = -\left[\mathcal{H}^{-1}S\right]_i - \frac{1}{2}\left[\mathcal{H}^{-1}E\right]_i.
\end{eqnarray*}
Utilizing the local landscape property of $\mathcal{H}$, we show in the supplementary material,
\begin{align}
    \left[\mathcal{H}^{-1}S\right]_i = \left(\sum_{j=1}^{p_2}\sum_{t=1}^T\partial_{\pi^2}l_{ijt}F_t^0c_j^0c_j^{0\prime}F_t^{0\prime}\right)^{-1}\sum_{j=1}^{p_2}\sum_{t=1}^T\partial_{\pi}l_{ijt}F_t^0c_j^0 + O_p\left((p_1p_2T)^{-1/2}\right).\label{EQ:H-1S}
\end{align}
The intuition behind (\ref{EQ:H-1S}) is that \( \mathcal{H} \) is approximately block diagonal with the aid of the augmented Lagrangian terms $P_1(\theta)$, $P_2(\theta)$ and also $P_3(\theta)$.

The elements within $\mathcal{H}$'s diagonal blocks are significantly larger than those in its off-diagonal blocks.
This structure of \( \mathcal{H} \) allows us to demonstrate that, in the expansion of \([\mathcal{H}^{-1}S]_i\), the additional terms resulting from the nonzero off-diagonal blocks collectively have an order of $O_p\left((p_1p_2T)^{-1/2}\right)$.
Then combing with Theorem 1, we show in the supplement that
\begin{align}
    \left\Vert\left[\mathcal{H}^{-1}E\right]_i \right\Vert= O_p\left(\frac{(p_1p_2T)^{3/\xi}}{L_{p_1p_2T}^2}\right).\label{EQ:H-1E}
\end{align}
Thus if \( \frac{(p_2T)^{1/2}(p_1p_2T)^{3/\xi}}{L_{p_1p_2T}^2} \rightarrow 0 \), \(\left\| [H^{-1}E]_i \right\| \) would be \( o_p((p_2T)^{-1/2}) \) and hence dominated by the first term on the right-hand side of (\ref{EQ:H-1S}).

    \begin{proposition}
        (Individual Rate). Under Assumptions \ref{ass1}-\ref{ass4}, and Assumption \ref{ass6},
        \begin{eqnarray*}
        \Vert\widehat{{r}}_i-{r}_i^0\Vert &=& O_p(L_{p_1p_2T}^{-1}) \mbox{ for each } i=1,\cdots,p_1,\\
        \Vert\widehat{f}_t-f_t^0\Vert &=& O_p(L_{p_1p_2T}^{-1}) \mbox{ for each } t=1,\cdots,T,\\
        \Vert\widehat{{c}}_j-{c}_j^0\Vert &= & O_p(L_{p_1p_2T}^{-1})  \mbox{ for each }j=1,\cdots,p_2.
\end{eqnarray*}
    \end{proposition}
    From Proposition 4, we see that the convergence rate of each component of ${\hat{\theta}}$ is the same as that of the least square estimator in LMFM by \cite{he2023Iterative}, but slower than that of the PE estimator in \cite{yu2022projected}, except that one of $\left\{(Tp_1)^{-1},(Tp_2)^{-1},(p_1p_2)^{-1}\right\}$ dominates the others. This is because both the maximum likelihood and the least square estimation derive estimators of ${\theta}$ jointly, while the PE method estimates $R$, $C$ and $F_t$ individually. Though the rate $O_p(L_{p_1p_2T}^{-1})$ is not sharp, but enough for deriving the order of $\left[\mathcal{H}^{-1}R\right]_i$. Then we have the following central limit theorem.

    \begin{theorem}\label{th2}
        (Individual Limit Distribution). Under Assumptions \ref{ass1}-\ref{ass6},
        \begin{align*}
            &\sqrt{p_2T}\left(\widehat{{r}}_i - {r}_{i}^0\right)\xrightarrow{d} \mathcal{N}\left(0,\Sigma_{iR}^{-1}\Omega_{iR}\Sigma_{iR}^{\prime-1}\right) \text{ if } \frac{\sqrt{p_2T}}{L^2_{p_1p_2T}}(p_1p_2T)^{3/\xi}\to 0, \\
            &\sqrt{p_1T}\left(\widehat{{c}}_j - {c}_{j}^0\right)\xrightarrow{d} \mathcal{N}\left(0,\Sigma_{jC}^{-1}\Omega_{jC}\Sigma_{jC}^{\prime-1}\right)  \text{ if } \frac{\sqrt{p_1T}}{L^2_{p_1p_2T}}(p_1p_2T)^{3/\xi}\to 0, \\
            &\sqrt{p_1p_2}(\widehat{{f}}_t - {f}_{t}^0)\xrightarrow{d} \mathcal{N}\left(0,\Sigma_{tF}^{-1}\Omega_{tF}\Sigma_{tF}^{\prime-1}\right) \text{ if } \frac{\sqrt{p_1p_2}}{L^2_{p_1p_2T}}(p_1p_2T)^{3/\xi}\to 0,
        \end{align*}
        where $\Sigma_{iR}, \Sigma_{jC}, \Sigma_{tF}$, $\Omega_{iR}$, $\Omega_{jC}$ and $\Omega_{tF}$ are defined in Assumption \ref{ass5}.
            \end{theorem}

\begin{remark}\label{rem1}
        The asymptotic variances of $\hat{r}_i$, $\hat{f}_t$ and $\hat{c}_j$ can be consistently estimated, respectively, by
        \begin{align*}
            \widehat{\text{var}}_{r} = \,&p_2T\left(\sum_{j=1}^{p_2}\sum_{t=1}^T\partial_{\pi^2}l_{ijt}(\hat{\pi}_{ijt})\hat{F}_t\hat{c}_j\hat{c}_j^{\prime}\hat{F}_t^{\prime}\right)^{-1}\left(\sum_{j=1}^{p_2}\sum_{t=1}^T\left(\partial_{\pi}l_{ijt}(\hat{\pi}_{ijt})\right)^2\hat{F}_t\hat{c}_j\hat{c}_j'\hat{F}_t'\right)\\
            &\times\left(\sum_{j=1}^{p_2}\sum_{t=1}^T\partial_{\pi^2}l_{ijt}(\hat{\pi}_{ijt})\hat{F}_t\hat{c}_j\hat{c}_j^{\prime}\hat{F}_t^{\prime}\right)^{-1},\\
           \widehat{ \text{var}}_{f} = \,&p_1p_2\left(\sum_{i=1}^{p_1}\sum_{j=1}^{p_2}\partial_{\pi^2}l_{ijt}(\hat{\pi}_{ijt})\hat{c}_j\otimes \hat{r}_i\hat{c}'_j\otimes \hat{r}'_i\right)^{-1}\left(\sum_{i=1}^{p_1}\sum_{j=1}^{p_2}\left(\partial_{\pi}l_{ijt}(\hat{\pi}_{ijt})\right)^2\hat{c}_j\otimes \hat{r}_i\hat{c}'_j\otimes \hat{r}'_i\right)\\
            &\times\left(\sum_{i=1}^{p_1}\sum_{j=1}^{p_2}\partial_{\pi^2}l_{ijt}(\hat{\pi}_{ijt})\hat{c}_j\otimes \hat{r}_i\hat{c}'_j\otimes \hat{r}'_i\right)^{-1},\\
           \widehat{ \text{var}}_{c} = \,&p_1T\left(\sum_{i=1}^{p_1}\sum_{t=1}^T\partial_{\pi^2}l_{ijt}(\hat{\pi}_{ijt})\hat{F}'_t\hat{r}_i\hat{r}_i^{\prime}\hat{F}_t\right)^{-1}\left(\sum_{i=1}^{p_1}\sum_{t=1}^T\left(\partial_{\pi}l_{ijt}(\hat{\pi}_{ijt})\right)^2\hat{F}'_t\hat{r}_i\hat{r}_i^{\prime}\hat{F}_t\right)\\
            &\times\left(\sum_{i=1}^{p_1}\sum_{t=1}^T\partial_{\pi^2}l_{ijt}(\hat{\pi}_{ijt})\hat{F}'_t\hat{r}_i\hat{r}_i^{\prime}\hat{F}_t\right)^{-1}.
        \end{align*}
\end{remark}

Theorem 2 generalizes
Theorem 3.2 of \cite{yu2022projected} allowing factors to be extracted from discrete or some other nonlinear models. And it allows the probability function to differ across rows, columns and samples. Thus the vast quantity of discrete data available in macroeconomic and financial studies can be effectively utilized, either by themselves or in combination with continuous data. An immediate consequence of Theorem \ref{th2} and Remark \ref{rem1} is the following corollary.
\begin{corollary}\label{cor1}
Under Assumptions \ref{ass1}-\ref{ass6},
        \begin{eqnarray*}
            &\sqrt{p_2T}\left(\widehat{{r}}_i - {r}_{i}^0\right)/\sqrt{\widehat{\text{var}}_{r}}\xrightarrow{d} \mathcal{N}(0, 1) \text{ if } \frac{\sqrt{p_2T}}{L^2_{p_1p_2T}}(p_1p_2T)^{3/\xi}\to 0, \\
            &\sqrt{p_1T}\left(\widehat{{c}}_j - {c}_{j}^0\right)/\sqrt{\widehat{\text{var}}_{c}}\xrightarrow{d} \mathcal{N}(0,1)  \text{ if } \frac{\sqrt{p_1T}}{L^2_{p_1p_2T}}(p_1p_2T)^{3/\xi}\to 0, \\
            &\sqrt{p_1p_2}\left(\widehat{{f}}_t - {f}_{t}^0\right)/\sqrt{\widehat{\text{var}}_{f}}\xrightarrow{d} \mathcal{N}(0,1) \text{ if } \frac{\sqrt{p_1p_2}}{L^2_{p_1p_2T}}(p_1p_2T)^{3/\xi}\to 0.
        \end{eqnarray*}
\end{corollary}

\subsection{Consistency in determining the number of factors}

Under the linear matrix factor model, \cite{yu2022projected} proposed an eigenvalue ratio test to consistently estimate the number of factors. While the eigenvalue ratio test is based on the ordered eigenvalues of the covariance matrix of $ {X}_{t}$, it is not suitable to
describe the correlation of non-continuous random variables. In
this article, we use the idea of information criterion to select the number
of factors. The estimator of $(k_1, k_2)$ is similar to the penalized loss (PC) based estimator of \cite{Bai2002De}, but is adaptive to the matrix observation and the likelihood function. Given the number of factors $l_1,l_2$, let $\widehat{{\theta}}^{(l_1,l_2)}$ be the estimator of ${\theta}$. Our PC-based estimator of $(k_1, k_2)$ is defined as
\begin{align}\label{EQ:k1k2select}
    (\widehat{k}_1,\widehat{k}_2) = \arg\min_{(l_1,l_2)}\left\{-\frac{1}{p_1p_2T}L\left(\widehat{{\theta}}^{(l_1,l_2)}\right)+(l_1+l_2)g(p_1,p_2,T)\right\},
\end{align}
where $g(p_1,p_2,T)$ is a prespecified penalty function. Theorem 3 below demonstrates that $(\widehat{k}_1,\widehat{k}_2)$ are consistent to $(k_1,k_2)$ by choosing a suitable $g(p_1,p_2,T)$.
    \begin{theorem}\label{th3}
        If $0<g(p_1,p_2,T)\to 0$, $g(p_1,p_2,T)L_{p_1p_2T}^2\to\infty$, under Assumptions \ref{ass1}-\ref{ass4}, and Assumption \ref{ass6}, $P\left(\widehat{k}_1 = k_1\right) \to 1$ and $P\left(\widehat{k}_2 = k_2\right) \to 1$, as $p_1,p_2,T\to\infty$.
    \end{theorem}

In our simulation studies and real data analysis, we
choose the penalty $g(p_1,p_2,T) = \frac{p_1+p_2+T}{p_1p_2T}\text{ln}\left(\frac{p_1p_2T}{p_1+p_2+T}\right)
$, which satisfies the
conditions in Theorem \ref{th3} and is confirmed to work well.

\subsection{Algorithms} \label{computation}

We shall introduce two algorithms, the two-stage alternating maximization and the minorization maximization to numerically calculate the maximum likelihood estimator. The second one is computationally simpler, but so far it has been shown to be only applicable to Probit, Logit and Tobit models (\cite{WANG2022180}).

\subsubsection{Two-stage alternating maximization (TSAM)}

Let $\mathcal{S}_{1j}$ be an indicator set pertaining to a specific variable type in the $j$-th column of $\mathbf{X}$ with respect to $(i,t)$, such that the cardinality of $S_{1j}$ and $p_1T$
are of comparable magnitude. Similarly, Let $\mathcal{S}_{2i}$ be an indicator set pertaining to a specific variable type in the $i$-th row of $\mathbf{X}$ with respect to $(j,t)$, such that the number of elements in $S_{2i}$
and $p_2T$ are comparable. And let $\mathcal{S}_{3t}$
be an indicator set associated with a particular variable type in the $t$-th sequence of $X$ relative to $(i,j)$, where the cardinality of $\mathcal{S}_{3t}$  and $p_1p_2$ are comparable in magnitude. Due to the finite number of variable types in $\mathbf{X}$, $S_{1i},S_{2j},S_{3t}$ are well defined.

\textbf{Algorithm 1}. Step 1 (Initialization): Randomly generate the initial values of $\widetilde{r}^{(0)}$ and $\widetilde{f}^{(0)}$.

Step 2 (Updating): For $k\geq 0$, calculate
\begin{align*}
    \widetilde{c}_j^{(k)} &= \mbox{argmax}_{c_j} \sum_{(i,t)\in \mathcal{S}_{1j}} \ l_{ijt}\left(\widetilde{r}_i^{(k)'}\widetilde{F}_t^{(k)}c_j\right), \, j=1,\ldots,p_2,\\
    \widetilde{r}_i^{(k+1)} &= \mbox{argmax}_{r_i} \sum_{(j,t)\in \mathcal{S}_{2i}} \ l_{ijt}\left({r}_i^{'}\widetilde{F}_t^{(k)}\widetilde{c}_j^{(k)}\right),\, i=1,\ldots,p_1,\\
     \widetilde{F}_t^{(k+1)} &= \mbox{argmax}_{F_t} \sum_{(i,j)\in \mathcal{S}_{3t}} \ l_{ijt}\left(\widetilde{r}_i^{(k+1)'}{F}_t\widetilde{c}_j^{(k)}\right),\, t=1,\ldots,T.
\end{align*}
Repeat the iteration until convergence and denote the derived estimators as $\widetilde{r}, \widetilde{f}$ and $\widetilde{c}$.

Step 3 (Correction): A second-stage update is then conducted based on the score function and Hessian matrix:
\begin{align}
    \widehat{{r}}_i &= \widetilde{{r}}_i - \left\{\partial_{{r}_ir_i'}L(X|\widetilde{{r}},\widetilde{{f}},\widetilde{{c}})\right\}^{-1}\partial_{{r}_i}L(X|\widetilde{{r}},\widetilde{{f}},\widetilde{{c}}), \ i=1,\ldots,p_1, \label{EQ:zrhat}\\
    \widehat{{f}}_t &= \widetilde{{f}}_t - \left\{\partial_{{f}_tf_t'}L(X|\widetilde{{r}},\widetilde{{f}},\widetilde{{c}})\right\}^{-1}\partial_{f_t}L(X|\widetilde{{r}},\widetilde{{f}},\widetilde{{c}}), \ t=1,\ldots,T,\label{EQ:zFhat}\\
      \widehat{{c}}_j &= \widetilde{{c}}_j - \left\{\partial_{{c}_jc_j'}L(X|\widetilde{{r}},\widetilde{{f}},\widetilde{{c}})\right\}^{-1}\partial_{{c_j}}L(X|\widetilde{{r}},\widetilde{{f}},\widetilde{{c}}), \ j=1,\ldots,p_2\label{EQ:zchat}.
\end{align}

Step 4 (Repetition): Repeat step 1-step 3 a number of times. Take the one with the largest likelihood.

Step 5 (Normalization): Let
$\widehat{R}^{(s)}$, $\widehat{C}^{(s)}$ and $\widehat{F}^{(s)}$ be the estimators from step 4.
To ensure the identification conditions in (\ref{constraint}), a normalization step could be applied to the factor and loading matrices. Specifically, do singular value decomposition to $\widehat{R}^{(s)}$ and $\widehat{C}^{(s)}$ as follows.
\begin{align*}
    \widehat {R}^{(s)} =  {U}_{R} {H}_R {V}_R :=  {U}_{R} {Q}_{R}, \quad\widehat{C}^{(s)} =  {U}_{C} {H}_C {V}_C :=  {U}_{C} {Q}_{C}.
\end{align*}
Define
\begin{align*}
     {\Sigma}_1 = \frac{1}{Tp_1p_2}\sum_{t=1}^T {Q}_{R} {F}_t {Q}_{C}' {Q}_{C} {F}_t' {Q}_{R}',  \quad  {\Sigma}_2 = \frac{1}{Tp_1p_2}\sum_{t=1}^T {Q}_{C} {F}_t' {Q}_{R}' {Q}_{R} {F}_t {Q}_{C}',
\end{align*}
and hence the eigenvalue decompositions
\begin{align*}
      {\Sigma}_1 =   {\Gamma}_1  {\Lambda}_1  {\Gamma}_1',  \quad  {\Sigma}_2 =   {\Gamma}_2  {\Lambda}_2  {\Gamma}_2'.
\end{align*}
Then, the normalized loading and factor score matrices are, respectively,
\begin{align}\label{EQ:Normalization1}
    \widehat {{R}} = \sqrt{p_1} {U}_R  {\Gamma}_1, \quad \widehat{{C}} = \sqrt{p_2} {U}_C  {\Gamma}_2 \ \ \mbox{and} \ \ \widehat {{F}}_t = (p_1p_2)^{-1/2}  {\Gamma}_1' {Q}_{R} {F}_t {Q}_{C}'  {\Gamma}_2.
\end{align}

\cite{Liu2023Gene} first proposed this algorithm for the generalized vector factor model, and the convergence of step 2 to a local maximum is given in their Proposition 2. To search for the global maximum, a common practice is to randomly choose the initial values multiple times and select the one with the highest likelihood among all local maxima. We generalize this approach to the matrix factor model setting. Although the computation for $(\widetilde{{r}},\widetilde{{f}},\widetilde{{c}})$ is simple, the efficiency may be lost since $\widetilde{{r}},\widetilde{{f}}$ and $\widetilde{{f}}$ are not obtained based on the log-likelihood function $L(X|r,f,c)$. To improve the efficiency, we then conduct a second-stage update in step 3, and the increase in efficiency from this one-step correction is validated in Section \ref{correction}. The strength of this algorithm lies in its ability to perform estimations in parallel across all rows, columns and sequences, leveraging existing packages. This ensures straightforward programming and computation, enhancing efficiency and convenience.

\subsubsection{ Minorization maximization (MM)}

\textbf{Algorithm 2}. Step 1 (Initialization): Randomly generate the initial values of $\widehat{r}^{(0)}$, $\widehat{c}^{(0)}$ and $\widehat{f}^{(0)}$.

Step 2 (Updating): For $k\geq 0$, first calculate $\widehat{x}_{ijt}^{(k)} = \widehat{r}_i^{(k)\prime}\widehat{F}_t^{(k)}\widehat{c}_j^{(k)} + \frac{1}{b_U}\partial_{\pi}l_{ijt}\left(\widehat{r}_i^{(k)\prime}\widehat{F}_t^{(k)}\widehat{c}_j^{(k)}\right)$ for $i=1,\ldots,p_1$, $j=1,\ldots,p_2$, $t=1,\ldots,T$, then update as follows.
$$
\left(\widehat{r}^{(k+1)}, \widehat{f}^{(k+1)}, \widehat{c}^{(k+1)}\right) = \argmin\sum_{i=1}^{p_1}\sum_{j=1}^{p_2}\sum_{t=1}^{T}\left(\hat{x}^{(k)}_{ijt}-{r}_i'{F}_t{c}_j\right)^2. $$
Iterate until $L\left(X| \widehat{r}^{(k+1)},\widehat{f}^{(k+1)},\widehat{c}^{(k+1)}\right) - L\left(X| \widehat{r}^{(k)},\widehat{f}^{(k)},\widehat{c}^{(k)}\right)\leq$ error, where error is the level of numerical tolerance.

Step 3 (Repetition): Repeat step 1 and step 2 a number of times. Take the one with the largest likelihood.

Step 4 (Normalization): Similar to Algorithm 1, the solution in Step 3 is normalized by (\ref{EQ:Normalization1}).

   The Minorization Maximization (MM) algorithm does not require alternation. Instead, it only necessitates the matrix factorization in step 2, which can be performed quickly using standard software packages. \cite{WANG2022180} applies this algorithm  to the generalized vector factor model and provides the proof of convergence for step 2 in their Appendix B. We are here extending it to the matrix factor models.

\section{Numerical Studies}\label{simulation}

In Sections \ref{setting} and \ref{compare}, we conduct simulation studies to assess the finite-sampling performance of the proposed method
(GMFM) by comparing it with the linear matrix factor model (LMFM). The accuracy between the factor loadings $\widehat{{R}}$ and ${R}^0$, evaluated by the smallest nonzero canonical
correlation between them, denoted by ccor$(\widehat{{R}}, {R}^0)$. Similarly, we also compute the smallest nonzero canonical correlation between $\widehat{{C}}$ and ${C}^0$, denoted by ccor$(\widehat{{C}}, {C}^0)$.
Canonical correlation has been widely used to measure the performance in factor analysis; see for example \cite{GOYAL2008252}, \cite{Doz2012}, \cite{bai2012statistical}, \cite{Liu2023Gene}). In sections \ref{number} and \ref{asympototic}, we examine the performance of the information criterion for selecting the number of factors and verify the asymptotic normality in Theorem \ref{th2}, separately. Section \ref{correction} investigates the efficiency gain of the second-stage correction in TSAM.

\subsection{Simulation Setting}\label{setting}

We generate the factor matrix sequence by an AR(1) model as $f_t = 0.2f_{t-1} + 0.2\epsilon_t$ where ${\epsilon_t}$'s are all generated from i.i.d. $\mathcal{N}(0_{k_1k_2},\mathbb{I}_{k_1k_2})$, $t=1,\cdots,T$. We draw the entries of ${R}$ and ${C}$ independently from the uniform distribution $U(0,1)$. We consider six settings with different variables and different combinations of $p_1 = 20,30,50$, $p_2 = 20,30,50$ and $T = 30,50$.

\noindent {\bf Case 1 (Gaussian variables with homoscedasticity)}:
$k_1 = k_2 = 2$, and  $x_{ijt}\sim \mathcal{N}({r}_i'{F}_t{c}_j,1)$.

\noindent {\bf Case 2 (Gaussian variables with heteroscedasticity)}: $k_1 = 1$, $k_2 = 3$, $x_{ijt} \sim \mathcal{N}({r}_i'{F}_t{c}_j,\tau_j^2)$ with $\tau_j = 0.1+2U_j$ and $U_j$'s are i.i.d from $U(0,1)$.

\noindent {\bf Case 3 (Poisson variables)}:
$k_1 = k_2 = 3$ and $x_{ijt} \sim \text{Poisson} (\exp{({r}_i'{F}_t{c}_j}))$.

\noindent {\bf Case 4 (The mixture of binary and count variables)}:
$k_1 = k_2 = 4$. For $i=1,\cdots,p_1$, $j = 1,\cdots,[p_2/2]$, $x_{ijt} \sim \text{Poisson} (\exp{({r}_i'{F}_t{c}_j}))$; for $j = [p_2/2]+1,\cdots,p_2$, $x_{ijt} \sim \text{Bernoulli} (1/(1+\exp{({r}_i'{F}_t{c}_j})))$.

\noindent {\bf Case 5 (The mixture of continuous and count variables)}:
$k_1 = k_2 = 5$. For $i=1,\cdots,p_1$, $j = 1,\cdots,[p_2/2]$, $x_{ijt} \sim \mathcal{N}({r}_i'{F}_t{c}_j,1)$; for $j = [p_2/2]+1,\cdots,p_2$, $x_{ijt} \sim \text{Poisson} (\exp{({r}_i'{F}_t{c}_j}))$.

\noindent {\bf Case 6 (The mixture of continuous, count and binary variables)}:
$k_1 = k_2 = 6$. For $i=1,\cdots,[p_1/2]$, $j = 1,\cdots,[p_2/2]$, $x_{ijt} \sim \mathcal{N}({r}_i'{F}_t{c}_j,1)$; for $i=1,\cdots,[p_1/2]$, $j = [p_2/2]+1,\cdots,p_2$, $x_{ijt} \sim \text{Poisson} (\exp{({r}_i'{F}_t{c}_j})$;  for $i=[p_1/2]+1,\cdots,p_1$, $j = 1,\cdots,[p_2/2]$, $x_{ijt} \sim \text{Poisson} (\exp{({r}_i'{F}_t{c}_j})$; for $i=[p_1/2]+1,\cdots,p_1$, $j = [p_2/2]+1,\cdots,p_2$, $x_{ijt} \sim \text{Bernoulli} (1/(1+\exp{({r}_i'{F}_t{c}_j}))$.

\subsection{Comparison with the Linear Factor Models} \label{compare}

Tables \ref{Tab:case123}-\ref{Tab:case4-6} report the average of ccor$(\widehat{{R}}, {R}^0)$ and ccor$(\widehat{{C}}, {C}^0)$ using the LMFM and GMFM based on 500 repetitions. The loadings and factors of GMFM are estimated and computed by our proposed TSAM algorithm, while the loadings and factors of LMFM are estimated by $\alpha$-PCA($\alpha=0$) in \cite{Chen2023Statistical}.
Theoretically, under conditions of homoscedasticity, the LMFM and the GMFM are equivalent. This equivalence is reflected in the comparable results observed for Case 1 in Table \ref{Tab:case123}. With heteroscedasticity, the GMFM demonstrates superior performance compared to the LMFM in case 2, as evident in Table \ref{Tab:case123}. Moreover, it is evident that the precision of $\widehat{{r}}$ and $\widehat{{C}}$ increases as $p_1$, $p_2$ and/or $T$ increases, which is consistent with our theoretical results in Theorems \ref{th1} and \ref{th2}.

In case 3 of table \ref{Tab:case123}, the performance of GMFM is significantly better than that of LMFM for the Poisson variables. Evidently, the LMFM is only effective for continuous variables, but performs poorly for discrete variables, and the estimation accuracy does not improve with increasing dimensionality. On the contrary, GMFM has a higher canonical correlation in estimating loadings for Poisson variables, showing a comparable accuracy to that of continuous variables.

Table \ref{Tab:case4-6} shows the result in Cases 4-6, where $ {X}_{t},t=1,\cdots,T$ are mixed variables
including Poisson, binary, and normal variables. It is clear that the LMFM does not work on mixed variables containing discrete variables, while the GMFM remains reliable in Table \ref{Tab:case4-6}. Particularly,
the canonical correlation of GMFM for both $\widehat{{R}}$ and $\widehat{{C}}$ increases as $p_1$, $p_2$ and/or $T$ increases, but that of LMFM does not.

\begin{table}[htbp]
\caption{The results for single-type variables.}
\label{Tab:case123}
\begin{center}
\renewcommand\arraystretch{0.5}
\begin{tabular}{ccccccccc}
\hline
\multirow{2}{*}{} &\multirow{2}{*}{}&\multirow{2}{*}{}  &\multicolumn{3}{c}{GMFM} &\multicolumn{3}{c}{LMFM}\\
\cmidrule(r){4-6}\cmidrule(r){7-9}
\multirow{2}{*}{} &T&$p_1 \backslash p_2$  &{$20$} &{$30$}  &{$50$} &{$20$} &{$30$}  &{$50$}\\\hline
& & & \multicolumn{6}{c}{Case 1 for Gaussian variables with homoscedasticity}\\\cmidrule(r){4-9}
\multirow{6}{*}{ccor$\left(\widehat{{R}}, {R}^0\right)$} &\multirow{3}{*}{$30$}
&$20$   &0.9520&0.9677 &0.9827 &0.8792 &09201 &0.9509\\
& &$30$  &0.9527&0.9690 &0.9836 &0.8882  &0.9248 &0.9590\\
& &$50$  &0.9608&0.9718 &0.9850 &0.9316  &0.9497 & 0.9724\\
&\multirow{3}{*}{$50$} &$20$  &0.9711&0.9833
&0.9879 &0.9204  &0.9551 &0.9725\\
& &$30$  &0.9734&0.9845 &0.9898 &0.9475  &0.9650 &0.9763\\
& &$50$  &0.9759&0.9845 &0.9909 &0.9590  &0.9728 &0.9840 \\
\\
\multirow{6}{*}{ccor$\left(\widehat{{C}}, {C}^0\right)$} &\multirow{3}{*}{$30$}
&$20$   &0.9389&0.9573 &0.9606 &0.8284 &0.8795 &0.9305\\
& &$30$  &0.9643&0.9669 &0.9702 &0.8779  &0.9092 &0.9427\\
& &$50$  &0.9788&0.9819 &0.9837 &0.9340  &0.9541 &0.9693 \\
&\multirow{3}{*}{$50$} &$20$  &0.9638&0.9718 &0.9744 &0.8973  &0.9381 &0.9563\\
& &$30$  &0.9790&0.9813 &0.9823 &0.9268  &0.9549 &0.9694\\
& &$50$  &0.9881&0.9898 &0.9903 &0.9592  & 0.9746& 0.9816\\\hline
& & & \multicolumn{6}{c}{Case 2 for Gaussian variables with heteroscedasticity}\\\cmidrule(r){4-9}
\multirow{6}{*}{ccor$\left(\widehat{{R}}, {R}^0\right)$} &\multirow{3}{*}{$30$}
&$20$   &0.9579&0.9771 &0.9874 &0.9341 &0.9566 &0.9740\\
& &$30$  &0.9650&0.9809 &0.9895 &0.9457  &0.9648 &0.9809\\
& &$50$  &0.9729 &0.9814 &0.9898 &0.9619  &0.9724 & 0.9844\\
&\multirow{3}{*}{$50$} &$20$  &0.9788&0.9881
&0.9933 &0.9618  &0.9765 &0.9859\\
& &$30$  &0.9832&0.9892 &0.9938 &0.9725  &0.9820 &0.9884\\
& &$50$  &0.9841&0.9905 &0.9941 &0.9775  &0.9859 &0.9912 \\
\\
\multirow{6}{*}{ccor$\left(\widehat{{C}}, {C}^0\right)$} &\multirow{3}{*}{$30$}
&$20$   &0.6849&0.7969 &0.8528 &0.1692 &0.1412 &0.2050\\
& &$30$  &0.7992&0.8821 &0.9094 &0.1881  &0.1559 &0.1996\\
& &$50$  &0.9151&0.9345 &0.9451 &0.1850  &0.1869 &0.2097 \\
&\multirow{3}{*}{$50$} &$20$  &0.8102 &0.8924 &0.9234 &0.1670  &0.1545 &0.2479\\
& &$30$  &0.9118  &0.9376 &0.9494 &0.1698  &0.1642 &0.2730\\
& &$50$  &0.9543  &0.9662 &0.9709 &0.1956  & 0.1907& 0.2650\\\hline
& & & \multicolumn{6}{c}{Case 3 for Possion variables}\\\cmidrule(r){4-9}
\multirow{6}{*}{ccor$\left(\widehat{{R}}, {R}^0\right)$} &\multirow{3}{*}{$30$}
&$20$   &0.9663&0.9806 &0.9882 &0.3492 &0.3393 &0.3666\\
& &$30$  &0.9697&0.9814 &0.9890 &0.2998  &0.2521 &0.3063\\
& &$50$  &0.9720&0.9824 &0.9891 &0.2140  &0.2165 & 0.2506\\
&\multirow{3}{*}{$50$} &$20$  &0.9808&0.9885
&0.9931 &0.3477  &0.4054 &0.4485\\
& &$30$  &0.9827&0.9889 &0.9933 &0.3101 &0.2969 &0.2905\\
& &$50$  &0.9836&0.9899 &0.9941 &0.2341  &0.2528 &0.2172 \\
\\
\multirow{6}{*}{ccor$\left(\widehat{{C}}, {C}^0\right)$} &\multirow{3}{*}{$30$}
&$20$   &0.9504&0.9658 &0.9687 &0.2711 &0.2624 &0.2177\\
& &$30$  &0.9699&0.9755 &0.9806 &0.2950  &0.2836 &0.2325\\
& &$50$  &0.9852&0.9867 &0.9876 &0.2811 &0.2738 &0.1843 \\
&\multirow{3}{*}{$50$} &$20$  &0.9732&0.9815 &0.9830 &0.3105  &0.2304 &0.1961\\
& &$30$  &0.9830&0.9859 &0.9879 &0.3055  &0.1895 &0.1612\\
& &$50$  &0.9905&0.9927 &0.9934 &0.2963  & 0.2441& 0.2149\\\hline
\end{tabular}
\end{center}
\end{table}

\begin{table}[htbp]
\caption{The results for mixed-type variables.}
\label{Tab:case4-6}
\begin{center}
\renewcommand\arraystretch{0.5}
\begin{tabular}{ccccccccc}
\hline
\multirow{2}{*}{} &\multirow{2}{*}{}&\multirow{2}{*}{}  &\multicolumn{3}{c}{GMFM} &\multicolumn{3}{c}{LMFM}\\
\cmidrule(r){4-6}\cmidrule(r){7-9}
\multirow{2}{*}{} &T&$p_1 \backslash p_2$  &{$20$} &{$30$}  &{$50$} &{$20$} &{$30$}  &{$50$}\\\hline
& & & \multicolumn{6}{c}{Case 4 for binary and count variables}\\\cmidrule(r){4-9}
\multirow{6}{*}{ccor$\left(\widehat{{R}}, {R}^0\right)$} &\multirow{3}{*}{$30$}
&$20$   &0.9408&0.9666 &0.9837 &0.3900 &0.3485 &0.3704\\
& &$30$  &0.9545&0.9718 &0.9857 &0.2693  &0.2651 &0.2580\\
& &$50$  &0.9617&0.9767 &0.9868 &0.2708  &0.2386 & 0.1882\\
&\multirow{3}{*}{$50$} &$20$  &0.9678&0.9819
&0.9911 &0.3961  &0.4086 &0.4384\\
& &$30$  &0.9731&0.9847 &0.9918 &0.3098 &0.3183 &0.2968\\
& &$50$  &0.9779 &0.9864 &0.9926 &0.2823  &0.2124 &0.2362 \\
\\
\multirow{6}{*}{ccor$\left(\widehat{{C}}, {C}^0\right)$} &\multirow{3}{*}{$30$}
&$20$   &0.7785&0.8926 &0.9207 &0.2160 &0.1705 &0.1533\\
& &$30$  &0.9162 &0.9350 &0.9500 &0.2126  &0.1508 &0.1405\\
& &$50$  &0.9515 &0.9627 &0.9717 &0.2079 &0.1756 &0.1412 \\
&\multirow{3}{*}{$50$} &$20$  &0.8724&0.9382 &0.9551 &0.2194  &0.1636 &0.1041\\
& &$30$  &0.9407 &0.9604 &0.9702 &0.1982  &0.1746 &0.1517\\
& &$50$  &0.9659 &0.9787 &0.9825 &0.1941  & 0.1476& 0.1375\\\hline
& & & \multicolumn{6}{c}{Case 5 for continuous and count variables}\\\cmidrule(r){4-9}
\multirow{6}{*}{ccor$\left(\widehat{{R}}, {R}^0\right)$} &\multirow{3}{*}{$30$}
&$20$   &0.9624&0.9818 &0.9913 &0.5255 &0.4910 &0.5678\\
& &$30$  &0.9705 &0.9843 &0.9917 &0.3885  &0.4213 &0.4085\\
& &$50$  &0.9791 &0.9868 &0.9929 &0.2685  &0.2784 & 0.2809\\
&\multirow{3}{*}{$50$} &$20$  &0.9814 &0.9898
&0.9946 &0.5002  &0.5464 &0.6054\\
& &$30$  &0.9849 &0.9916 &0.9954 &0.4544 &0.4507 &0.4529\\
& &$50$  &0.9875 &0.9926 &0.9961 &0.3168  &0.3032 &0.3323 \\
\\
\multirow{6}{*}{ccor$\left(\widehat{{C}}, {C}^0\right)$} &\multirow{3}{*}{$30$}
&$20$   &0.8882&0.9675 &0.9766 &0.2878 &0.2077&0.1607\\
& &$30$  &0.9596 &0.9785 &0.9850 &0.2477  &0.2497 &0.1767\\
& &$50$  &0.9843 &0.9895 &0.9912 &0.3200 &0.1805 &0.1495 \\
&\multirow{3}{*}{$50$} &$20$  &0.9473&0.9806 &0.9867 &0.2900 &0.2060 &0.1764\\
& &$30$  &0.9745 &0.9875 &0.9914 &0.2574  &0.2566 &0.1795\\
& &$50$  &0.9902 &0.9937 &0.9950 &0.3306  & 0.1877& 0.1635\\\hline
& & & \multicolumn{6}{c}{Case 6 for continuous, count and binary variables }\\\cmidrule(r){4-9}
\multirow{6}{*}{ccor$\left(\widehat{{R}}, {R}^0\right)$} &\multirow{3}{*}{$30$}
&$20$   &0.6565 &0.8353 &0.9498 &0.2703 &0.3290 &0.2900\\
& &$30$  &0.8167 &0.9290 &0.9777 &0.2364  &0.2423 &0.2070\\
& &$50$  &0.9493 &0.9753 &0.9880 &0.1303  &0.1690 & 0.1637\\
&\multirow{3}{*}{$50$} &$20$  &0.7052 &0.8737
&0.9620 &0.3071  &0.3589 &0.3450\\
& &$30$  &0.8969 &0.9583 &0.9849 &0.1675 &0.2789 &0.2523\\
& &$50$  &0.9662 &0.9850 &0.9930 &0.1593  &0.1828 &0.1758\\
\\
\multirow{6}{*}{ccor$\left(\widehat{{C}}, {C}^0\right)$} &\multirow{3}{*}{$30$}
&$30$   &0.5365 &0.7318 &0.9250 &0.2667 &0.1670 &0.1246\\
& &$60$  &0.6170 &0.8871 &0.9621 &0.1935  &0.1428 &0.1224\\
& &$90$  &0.8769 &0.9643 &0.9848 &0.2093 &0.1545 &0.1241 \\
&\multirow{3}{*}{$50$} &$30$  &0.5823 &0.8080 &0.9544 &0.2584 &0.2362 &0.1142\\
& &$60$  &0.7168 &0.9095 &0.9768 &0.2004  &0.1887 &0.1074\\
& &$90$  &0.8829 &0.9742 &0.9896 &0.1775  & 0.1606 & 0.1457\\\hline
\end{tabular}
\end{center}
\end{table}
\subsection{Model Selection} \label{number}

This section aims to verify the performance of the proposed methods for
selecting the number of row and column factors.
Table \ref{Tab:k1k2} reports the average of $\widehat{k}_1$ and $\widehat{k}_2$ based on $100$ repetitions, where
the candidates of $k_1$ and $k_2$ are the integers from $1$ to $8$. In Table \ref{Tab:k1k2},
it can be seen that the PC-type information criterion (\ref{EQ:k1k2select}) works well in all six cases considered with different $k_1$ and $k_2$. We found that the LMFM using $\alpha$-PCA $(\alpha = 0)$
by \cite{chen2022factor} and the projected estimation (PE) by \cite{yu2022projected} both failed under the PC information criterion, which tends to choose $(1, 1)$ in all the six cases.

\begin{table}[htbp]\label{table3}
\caption{The average of the estimated number of row and column factors by using our proposed information criterion based on GMFM from $100$ repetitions.}
\label{Tab:k1k2}
\begin{center}
\renewcommand\arraystretch{0.8}
\resizebox{\linewidth}{!}{
\begin{tabular}{cccccccc}
\hline
T&$p_1 \backslash p_2$  &{$20$} &{$30$}  &{$50$} &{$20$} &{$30$}  &{$50$}\\\hline
 & & \multicolumn{3}{c}{Case 1: $(k_1,k_2) = (2,2)$} & \multicolumn{3}{c}{Case 2: $(k_1,k_2) = (1,3)$}\\\cmidrule(r){3-5}\cmidrule(r){6-8}
 \multirow{3}{*}{$30$}
&$20$   &$(1.00,1.00)$ &$(1.00,1.00)$ &$(1.30,1.30)$ &$(1.00,1.00)$&$(1.00,1.00)$ &$(1.00,1.17)$\\
 &$30$  &$(1.15,1.15)$ &$(1.60,1.60)$ &$(1.96,1.96)$ &$(1.00,1.00)$ &$(1.00,1.00)$ &$(1.00,1.30)$\\
 &$50$  &$(1.65,1.50)$ &$(1.85,1.60)$ &$(2.00,2.00)$ &$(1.00,1.00)$ &$(1.00,1.00)$ &$(1.00,2.02)$\\
 \multirow{3}{*}{$50$}
&$20$  &$(1.32,1.32)$ &$(1.76,1.76)$
&$(2.00,2.00)$ &$(1.20,1.10)$ &$(1.0,1.22)$
&$(1.00,1.50)$\\
 &$30$  &$(1.64,1.62)$ &$(1.96,1.94)$ &$(2.00,2.00)$ &$(1.00,1.10)$ &$(1.00,1.42)$ &$(1.00,2.00)$\\
 &$50$  &$(1.98,1.96)$ &$(2.00,2.00)$ &$(2.00,2.00)$  &$(1.00,1.20)$ &$(1.00,1.82)$ &$(1.00,3.00)$ \\
\hline
 & & \multicolumn{3}{c}{Case 3: $(k_1,k_2) = (3,3)$}& \multicolumn{3}{c}{Case 4: $(k_1,k_2) = (4,4)$}\\\cmidrule(r){3-5}\cmidrule(r){6-8}
 \multirow{3}{*}{$30$}
&$20$   &$(1.00,1.00)$&$(1.00,1.00)$ &$(1.00,1.00)$ &$(1.00,1.00)$ &$(1.00,1.00)$ &$(1.00,1.00)$\\
 &$30$  &$(1.00,1.00)$ &$(1.40,1.40)$ &$(2.15,2.05)$ &$(1.00,1.00)$ &$(2.12,2.04)$ &$(2.52,2.36)$\\
 &$50$  &$(1.20,1.10)$ &$(2.05,1.85)$ &$(2.96,2.54)$  &$(1.00,1.00)$ &$(2.40,2.20)$ &$(3.64,3.36)$ \\
 \multirow{3}{*}{$50$}
&$20$  &$(1.00,1.00)$ &$(1.30,1.25)$
&$(2.50,2.48)$   &$(1.00,1.00)$ &$(1.52,1.30)$
&$(2.60,2.60)$\\
 &$30$  &$(1.70,1.60)$ &$(2.60,2.35)$&$(3.00,2.93)$ &$(2.80,2.52)$ &$(3.36,3.04)$ &$(4.00,3.82)$\\
 &$50$  &$(2.32,2.16)$ &$(3.00,2.86)$ &$(3.00,3.00)$ &$(3.04,2.56)$ &$(3.95,3.40)$ &$(4.00,4.00)$\\
\hline
 & & \multicolumn{3}{c}{Case 5: $(k_1,k_2) = (5,5)$}& \multicolumn{3}{c}{Case 6: $(k_1,k_2) = (6,6)$}\\\cmidrule(r){3-5}\cmidrule(r){6-8}
 \multirow{3}{*}{$30$}
&$20$   &$(4.72,4.60)$&$(5.00,4.85)$ &$(5.00,5.00)$  &$(4.20,3.92)$&$(5.35,4.60)$
&$(5.46,5.86)$\\
 &$30$  &$(5.00,4.68)$ &$(5.00,5.00)$ &$(5.00,5.00)$ &$(5.52,4.40)$ &$(5.80,5.60)$ &$(5.86,6.00)$\\
 &$50$  &$(5.00,5.00)$ &$(5.00,5.00)$ &$(5.00,5.00)$  &$(5.82,5.30)$ &$(6.00,5.80)$ &$(6.00,6.00)$\\
 \multirow{3}{*}{$50$}
&$20$  &$(5.00,4.76)$ &$(5.00,5.00)$
&$(5.00,5.00)$ &$(5.30,4.45)$ &$(5.50,5.30)$
&$(5.80,6.00)$\\
 &$30$  &$(5.00,4.80)$ &$(5.00,5.00)$ &$(5.00,5.00)$ &$(5.70,4.90)$ &$(6.00,5.80)$&$(6.00,6.00)$ \\
 &$50$  &$(5.00,5.00)$ &$(5.00,5.00)$ &$(5.00,5.00)$ &$(6.00,5.12)$ &$(6.00,5.82)$ &$(6.00,6.00)$  \\
\hline
\end{tabular}}
\end{center}
\end{table}

\subsection{Asymptotic distribution}\label{asympototic}

In this section, we assess the adequacy of the asymptotic distributions in Theorem \ref{th2}. We consider the case with $k_1=k_2=1$ and generate $f_{t}$ from i.i.d.  $\mathcal{N}(0,1)$. The generation and normalization of $R$ and $C$ are the same as Section \ref{setting}. For the given $R$, $F$ and $C$, we consider three data generating processes (DGPs) for $X_{ijt}$.

\noindent {\bf DGP 1 (Logit)}:
 $X_{ijt}$ is a binary random variable and $P(X_{ijt} = 1) = \Psi(r_i'F_tc_j)$, where
$\Psi(z) = 1/\left(1+e^{-z}\right)$.

\noindent {\bf DGP 2 (Probit)}:
$X_{ijt}$ is a binary random variable and $P(X_{ijt} = 1) = \Phi(r_i'F_tc_j)$, where $\Phi(\cdot)$ is the cumulative distribution function of standard normal distribution.

\noindent {\bf DGP 3(Mixed)}:
 For $t=1,\ldots,T$, $i=1,\ldots,p_1$, $j=1,\ldots,p_2/3$, $X_{ijt}\sim N(r_i'F_tc_j,1)$; for $t=1,\ldots,T$, $i=1,\ldots,p_1$, $j=p_2/3+1,\ldots,2p_2/3$, $X_{ijt}$ is a binary random variable and $P(X_{ijt} = 1) = \Psi(r_i'F_tc_j)$; for $t=1,\ldots,T$, $i=1,\ldots,p_1$, $j=2p_2/3+1,\ldots,p_2$, $X_{ijt}$ is a binary random variable and $P(X_{ijt} = 1) = \Phi(r_i'F_tc_j)$.

\begin{figure}[htbp]
\centering
    \begin{minipage}[t]{0.30\linewidth}
        \centering
        \includegraphics[width=0.65\linewidth]{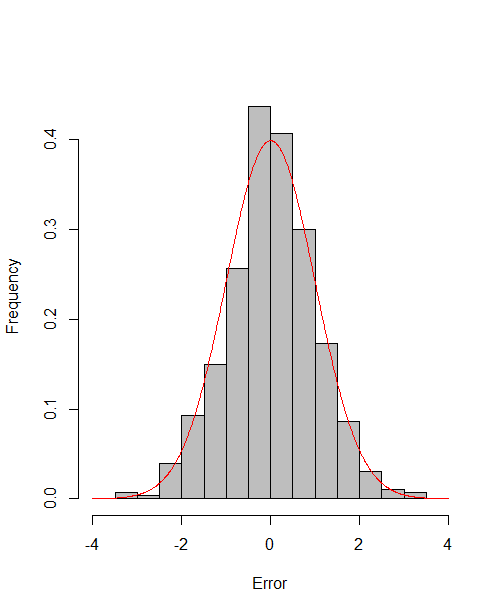}
        \caption*{GDP1 by MM}
    \end{minipage}
    \begin{minipage}[t]{0.30\linewidth}
        \centering
        \includegraphics[width=0.65\linewidth]{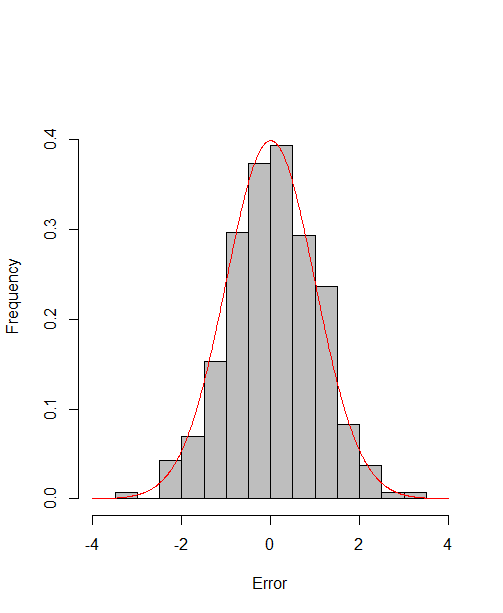}
        \caption*{GDP2 by MM}
    \end{minipage}
    \begin{minipage}[t]{0.30\linewidth}
        \centering
        \includegraphics[width=0.65\linewidth]{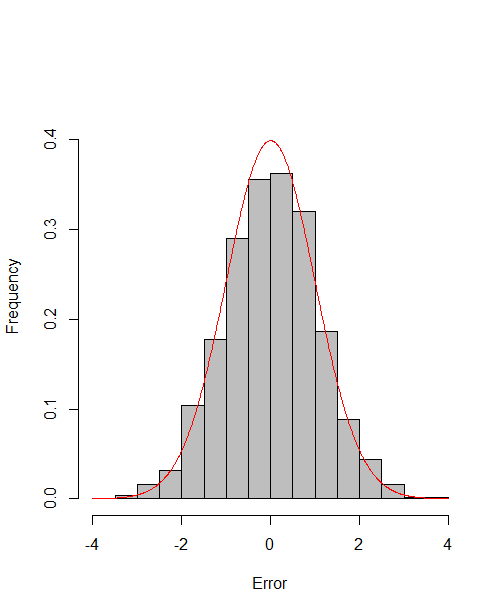}
        \caption*{GDP3 by MM}
    \end{minipage}
    \begin{minipage}[t]{0.30\linewidth}
        \centering
        \includegraphics[width=0.65\linewidth]{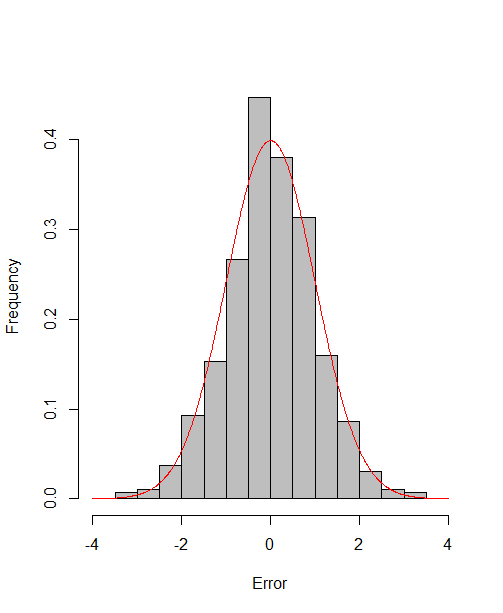}
        \caption*{GDP1 by TSAM}
    \end{minipage}
    \begin{minipage}[t]{0.30\linewidth}
        \centering
        \includegraphics[width=0.65\linewidth]{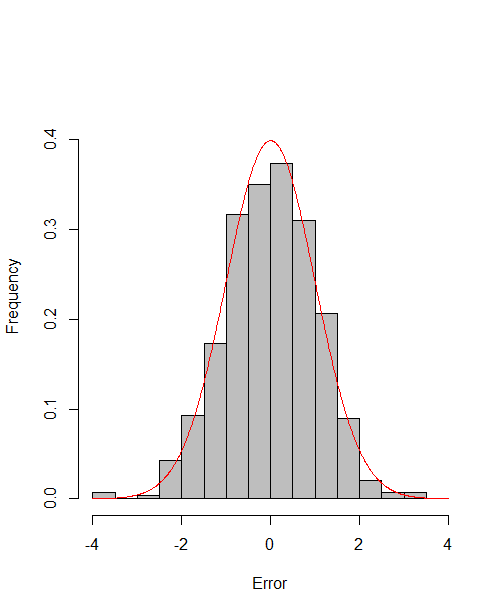}
        \caption*{GDP2 by TSAM}
    \end{minipage}
    \begin{minipage}[t]{0.30\linewidth}
        \centering
        \includegraphics[width=0.65\linewidth]{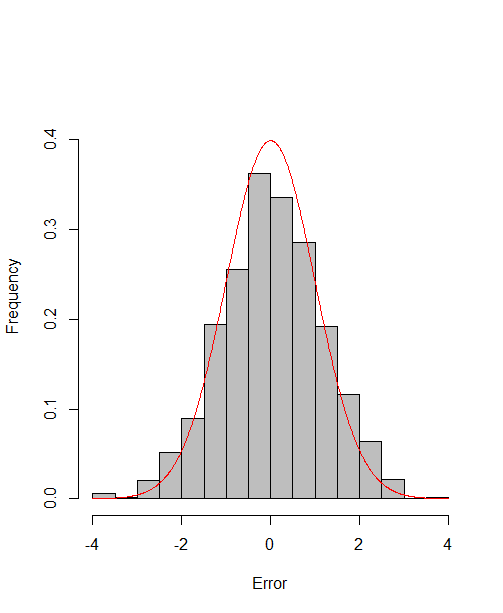}
        \caption*{GDP3 by TSAM}
    \end{minipage}
    \caption{Empirical distributions of $\hat{r}_{1}$ after standardization under various settings, over $1000$ replications with $T=p_1=p_2=50$. The red real lines are the density function of the standard normal
distribution.}\label{Fig:asyR}
    \end{figure}

\begin{figure}[htbp]
\centering
    \begin{minipage}[t]{0.30\linewidth}
        \centering
        \includegraphics[width=0.65\linewidth]{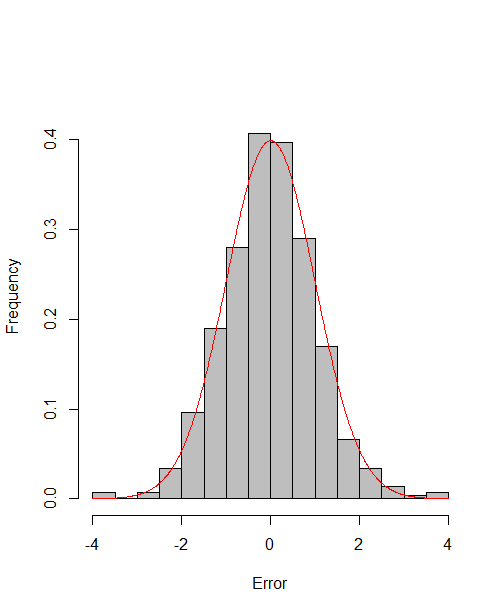}
        \caption*{GDP1 by MM}
    \end{minipage}
    \begin{minipage}[t]{0.3\linewidth}
        \centering
        \includegraphics[width=0.65\linewidth]{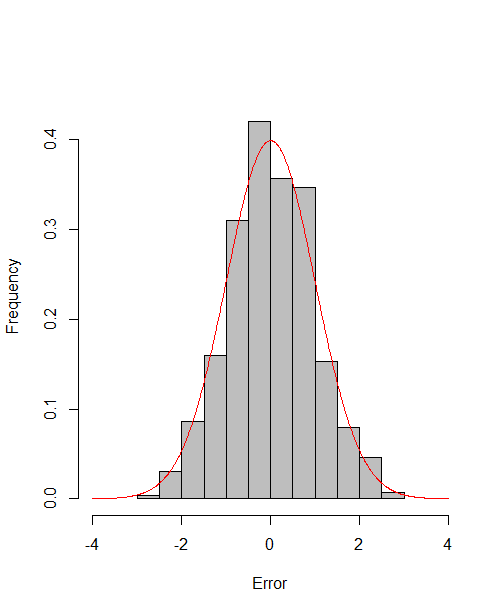}
        \caption*{GDP2 by MM}
    \end{minipage}
    \begin{minipage}[t]{0.3\linewidth}
        \centering
        \includegraphics[width=0.65\linewidth]{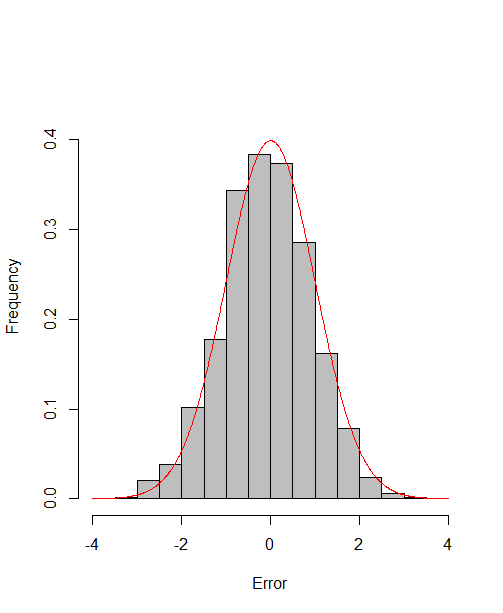}
        \caption*{GDP3 by MM}
    \end{minipage}
    \begin{minipage}[t]{0.3\linewidth}
        \centering
        \includegraphics[width=0.65\linewidth]{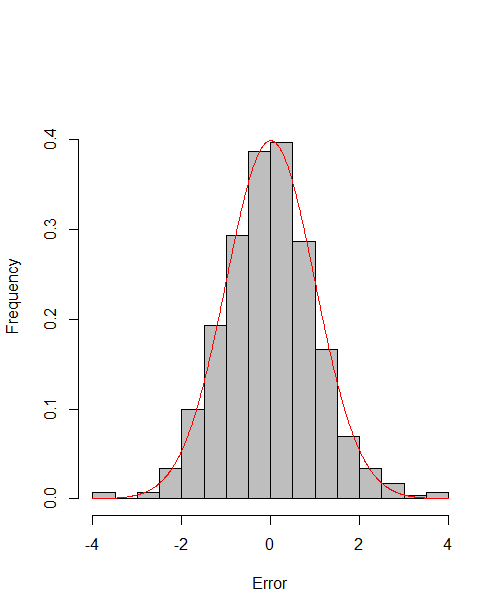}
        \caption*{GDP1 by TSAM}
    \end{minipage}
    \begin{minipage}[t]{0.30\linewidth}
        \centering
        \includegraphics[width=0.65\linewidth]{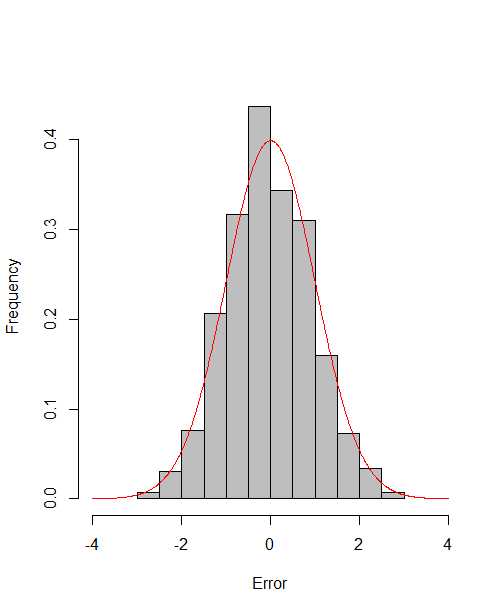}
        \caption*{GDP2 by TSAM}
    \end{minipage}
    \begin{minipage}[t]{0.30\linewidth}
        \centering
        \includegraphics[width=0.65\linewidth]{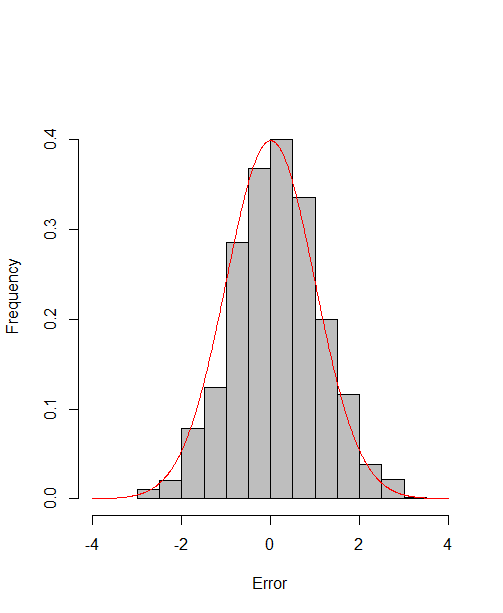}
        \caption*{GDP3 by TSAM}
    \end{minipage}
    \caption{Empirical distributions of $\hat{c}_{1}$ after standardization under various settings, over $1000$ replications with $T=p_1=p_2=50$. The red real lines are the density function of the standard normal
distribution.}\label{Fig:asyC}
    \end{figure}

\begin{figure}[htbp]
\centering
    \begin{minipage}[t]{0.30\linewidth}
        \centering
        \includegraphics[width=0.65\linewidth]{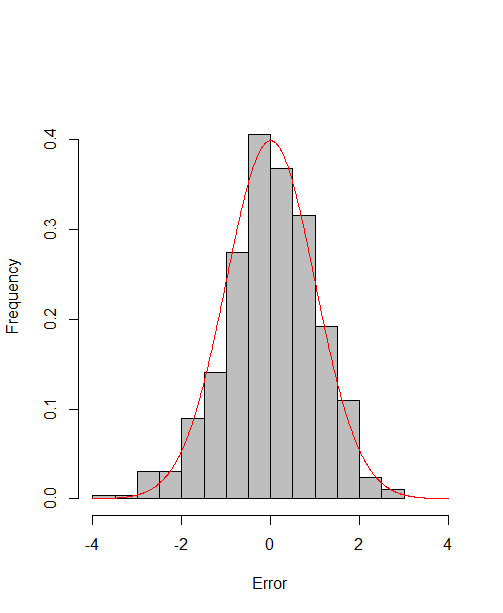}
        \caption*{GDP1 by MM}
    \end{minipage}
    \begin{minipage}[t]{0.30\linewidth}
        \centering
        \includegraphics[width=0.65\linewidth]{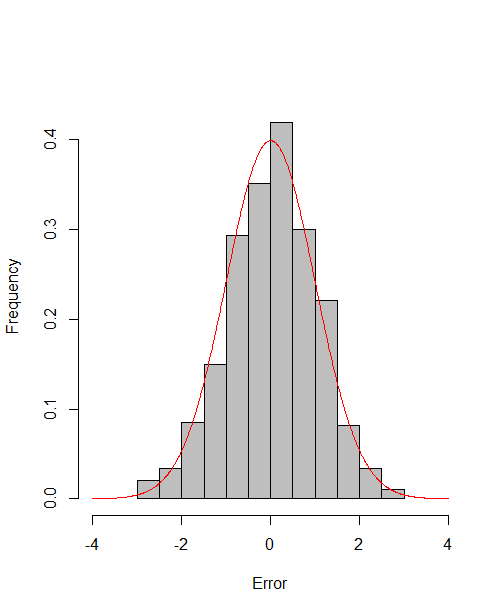}
        \caption*{GDP2 by MM}
    \end{minipage}
    \begin{minipage}[t]{0.30\linewidth}
        \centering
        \includegraphics[width=0.65\linewidth]{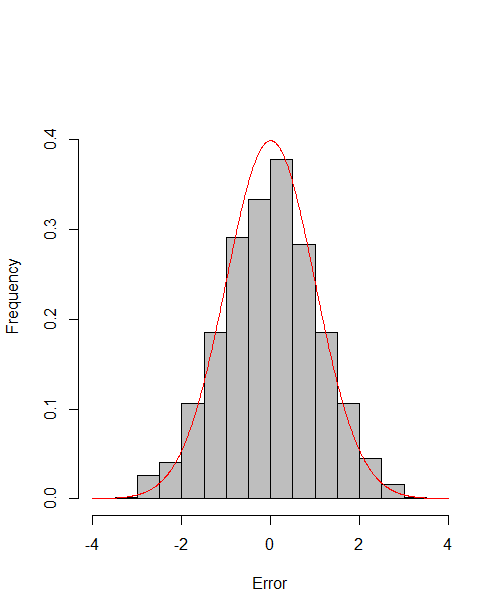}
        \caption*{GDP3 by MM}
    \end{minipage}
    \begin{minipage}[t]{0.30\linewidth}
        \centering
        \includegraphics[width=0.65\linewidth]{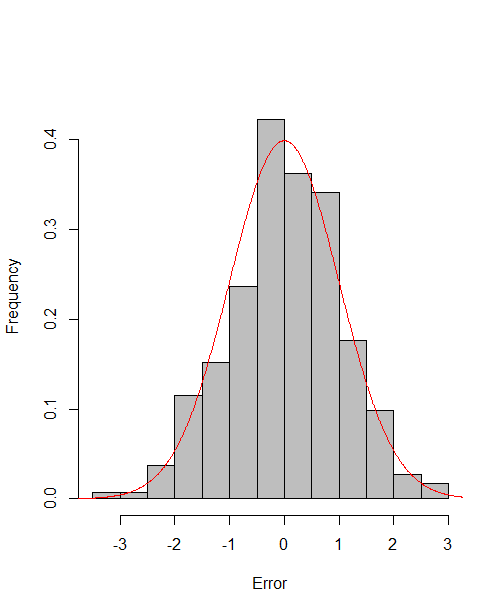}
        \caption*{GDP1 by TSAM}
    \end{minipage}
    \begin{minipage}[t]{0.30\linewidth}
        \centering
        \includegraphics[width=0.65\linewidth]{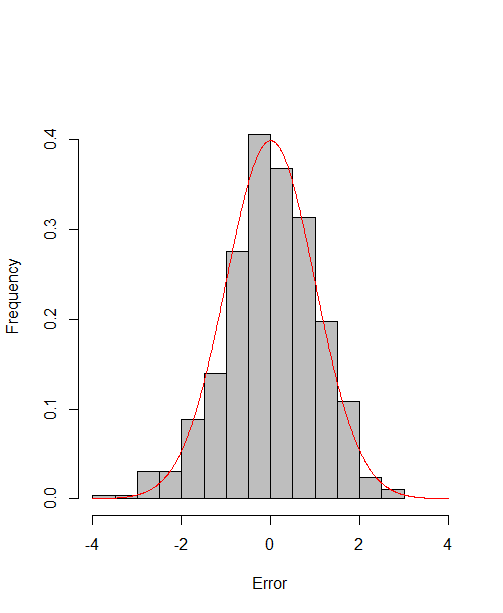}
        \caption*{GDP2 by TSAM}
    \end{minipage}
    \begin{minipage}[t]{0.30\linewidth}
        \centering
        \includegraphics[width=0.65\linewidth]{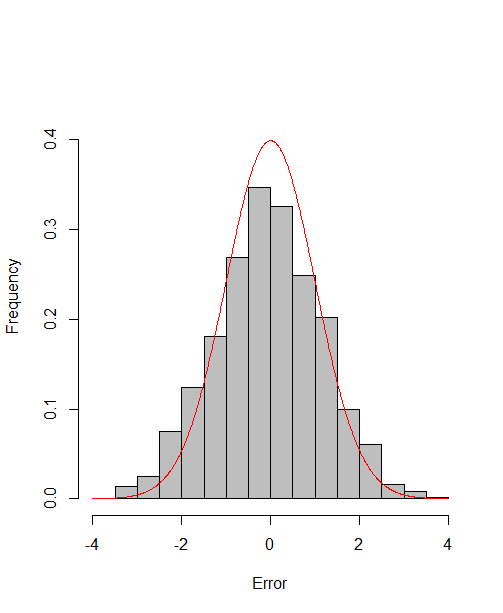}
        \caption*{GDP3 by TSAM}
    \end{minipage}
    \caption{Empirical distributions of $\hat{f}_{1}$ after standardization under various settings, over $1000$ replications with $T=p_1=p_2=50$. The red real lines are the density function of the standard normal
distribution.}\label{Fig:asyF}
    \end{figure}

Under such cases, by Theorem \ref{th2},
\begin{align*}
\sqrt{p_2T}(\hat{r}_{i} - r^0_{i})\xrightarrow{d} \mathcal{N}(0,p_2T\Sigma_{iR}^{-1}),\\
\sqrt{p_1T}(\hat{c}_{j} - c^0_{j})\xrightarrow{d} \mathcal{N}(0,p_1T\Sigma_{jC}^{-1}),\\
\sqrt{p_1p_2}(\hat{f}_{t} - f^0_{t})\xrightarrow{d} \mathcal{N}(0,p_1p_2\Sigma_{tF}^{-1}),
\end{align*}
which reach the asymptotic efficiency bounds.
Figure \ref{Fig:asyR}, \ref{Fig:asyC} and \ref{Fig:asyF} show the normalized histograms of $\Sigma_{1R}^{1/2}(\hat{r}_{1} - r^0_{1}), \Sigma_{1C}^{1/2}(\hat{c}_{1} - c^0_{1})$ and $\Sigma_{1F}^{1/2}(\hat{f}_{1} - f^0_{1})$ for DGP1-DGP3 by TSAM and MM separately, and the standard normal density curve is overlaid on them for comparison. As for the step 2 in MM, we choose $b_U = 1/4$
for the Logit case and $b_U = 1$ for the Probit case by the definition of $b_U$ in Section \ref{setup}. The standard normal density curve clearly provides a good approximation across all subfigures.

    \subsection{Efficiency Gain of the One-Step Correction in TSAM}\label{correction}

To demonstrate the efficiency improvement from the second-stage update in TSAM, we illustrate the results using Case 5 with $T=20$. In the first stage of TSAM, we obtain the estimator using continuous variables (S1N), count variables (S1P), and both (S1NP). As shown in Table \ref{Tab:effof5}, the performance of the first-stage estimators improves as $p_1$, $p_2$, and/or $T$ increase. Notably, S1NP demonstrates the best performance, while S1P and S1N exhibit similar levels of effectiveness. This aligns with our expectation since S1NP utilizes the most comprehensive information. The second-stage updating significantly enhances the performance of the estimators obtained from stage one, especially when the sample size is small.

\begin{table}[htbp]
\caption{The results of Case 5 with $T=20$. The estimators obtained in the first stage, which are based solely on Normal variables, solely on Poisson variables, and on a combination of both, are abbreviated as S1N, S1P and S1NP, respectively. The estimators arising from the second-stage update are designated as OSN, OSP and OSNP. The values of ccorR and ccorC represent the averaged ccor$(\widehat{{R}}, {R}^0)$ and ccor$(\widehat{{C}}, {C}^0)$, respectively.}
\label{Tab:effof5}
\begin{center}
\begin{tabular}{cccccccccc}
\hline
 &\multicolumn{3}{c}{$(p_1,p_2) = (30,20)$} &\multicolumn{3}{c}{$(p_1,p_2) = (30,30)$}&\multicolumn{3}{c}{$(p_1,p_2) = (30,50)$}\\
\hline
 &S1N&S1P   &S1NP &S1N&S1P   &S1NP &S1N&S1P   &S1NP\\
{ccorR} &0.9328&0.9388 &0.9702 &0.9502&0.9583 &0.9787
    &0.9703 &0.9776 &0.9879\\
{ccorC}&0.8930  &0.8612 & 0.9470&0.9055 &0.9533 &0.9652 &0.9699 &0.9695 &0.9764\\
\\
\multirow{2}{*}{} &OSN&OSP   &OSNP &OSN&OSP   &OSNP &OSN&OSP   &OSNP\\
{ccorR}&0.9692  &0.9701&0.9753 &0.9761 &0.9761 &0.9791  &0.9872&0.9877&0.9883\\
{ccorC}&0.9456  &0.9424 &0.9542 &0.9500 &0.9689
&0.9699 &0.9766&0.9766&0.9769\\
\hline
\end{tabular}%
\end{center}
\end{table}
\section{Real Data Analysis}\label{real data}
In this section, we analyze the operating performance of the listed high-tech manufacturing companies in China. These comprehensive datasets are sourced from various reliable resources, encompassing publicly accessible financial reports, corporate social responsibility statements, and stock exchange-released trading data. The data can be downloaded from https://data.csmar.com. This dataset encompasses 22 indicators for 12 manufacturing companies, spanning a total of 68 quarters from 2007-Q1 to 2023-Q2. It is worth noting that, unlike in the general LMFM literature (\cite{wang2019factor},\cite{chen2022factor},\cite{yu2022projected}), in addition to 18 continuous variables, there are four binary variables. These binary variables represent corporate social responsibility, such as the protection of employees' rights and interests, environmental and sustainable development initiatives, as well as public relations and social welfare undertakings. Indicator disclosure is marked as $1$, while non-disclosure is marked as $0$. Furthermore, the economic indicators of these companies can be categorized into four main groups: profitability, development ability, operational ability, and solvency. A detailed list of these indicators is presented in the supplementary material. The enterprises belong to three high-tech industries: pharmaceutical manufacturing, electronic equipment manufacturing, and transportation manufacturing.

The missing rate is low in this data set ($0.05\%$), and we use simple linear interpolation to fill in the missing values. Each original univariate time series is transformed by taking the second difference. We also standardize each of the transformed series to avoid the effects of non-zero mean or disparate variances.

\begin{figure}[htbp]
\centering
    \begin{minipage}[t]{0.48\linewidth}
        \centering
        \includegraphics[width=0.8\linewidth]{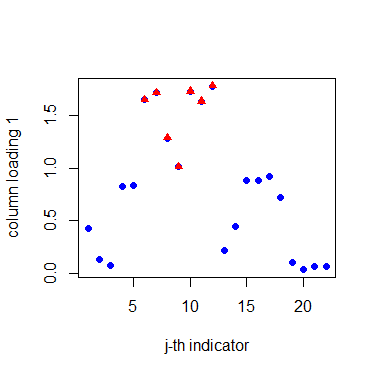}
    \end{minipage}
    \vspace{3mm}
    \begin{minipage}[t]{0.48\linewidth}
        \centering
        \includegraphics[width=0.8\linewidth]{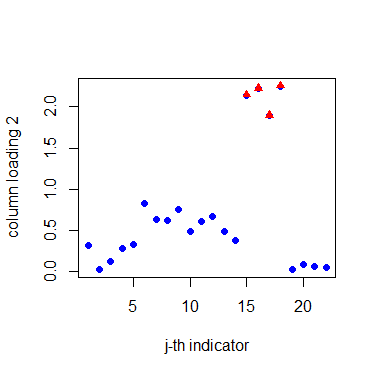}
    \end{minipage}
     \begin{minipage}[t]{0.48\linewidth}
        \centering
        \includegraphics[width=0.8\linewidth]{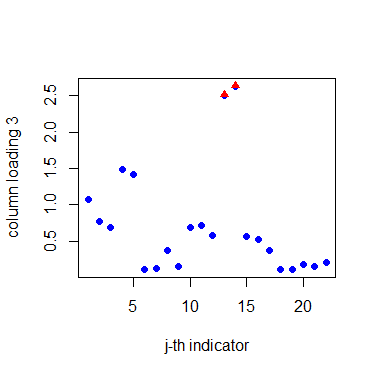}
    \end{minipage}
    \vspace{3mm}
    \begin{minipage}[t]{0.48\linewidth}
        \centering
        \includegraphics[width=0.8\linewidth]{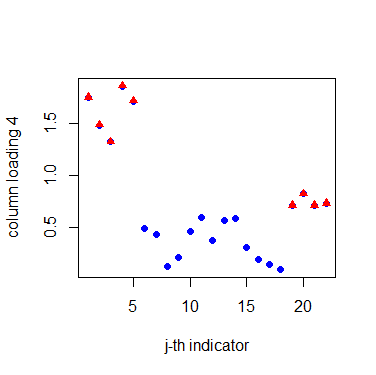}
    \end{minipage}
    \caption{The absolute values among the four column loadings, where the red triangular dots are the first several largest absolute values of each loading.}\label{Fig:real2}
    \end{figure}

Before the estimation of matrix factors and matrix loadings, we need to determine the numbers of row and column factors first. Our proposed criterion suggests taking $k_1=3$, $k_2 = 4$. For the row factors, Table \ref{Tab:real1} shows that they are closely related to the industry classification. Companies in the same industry often have similar loadings on the factors. These 12 companies are naturally divided into three groups, namely pharmaceutical manufacturing (P1, P2, P3, P4), electronic equipment manufacturing (E1, E2, E3, E4, E5), and transportation manufacturing (T1, T2, T3). For the column factors, Figure \ref{Fig:real2} divides the indicators into four groups: profitability (G1), operational ability (G2), solvency (G3), and development ability (G4). Notably, the disclosure of social responsibility has been incorporated into the indicators of development capability. A company with a strong sense of social responsibility generally tends to have better development prospects. This categorization aligns with economic explanations. The detailed grouping of these indicators is clearly outlined in the supplementary material.

\begin{table}[htbp]
\caption{Row (enterprises) loading matrices (multiplied by 10) by GMLE.}
\label{Tab:real1}
\begin{center}
\begin{tabular}{ccccccccccccc}
\hline
Row Factor&E1&E2&E3&E4&E5&P1&P2&P3&P4&T1&T2&T3\\
\hline
F1&-5&-3&2&0&-2&\cellcolor{Ocean}-21&\cellcolor{Ocean}-20&\cellcolor{Ocean}-16&\cellcolor{Ocean}-9&0&-2&1\\
F2&-3&-5&0&-8&-6&1&1&0&3&\cellcolor{Ocean}20&\cellcolor{Ocean}17&\cellcolor{Ocean}20\\
F3&\cellcolor{Ocean}-17&\cellcolor{Ocean}-13&\cellcolor{Ocean}-10&\cellcolor{Ocean}-20&\cellcolor{Ocean}-13&2&1&0&5&-5&-6&-8\\
\hline
\end{tabular}%
\end{center}
\end{table}

The canonical correlation
of ${R}$ (resp. ${C}$) for LMFM (estimating by $\alpha$-PCA with $\alpha = 0$) and GMFM is 0.76 ( resp. 0.70) respectively, showing that the loading estimators of LMFM are different from those of GMFM. In addition, we calculated the negative scaled log-likelihood $-2L/(Tp_1p_2)$ for GMFM and LMFM to measure the goodness of fit ($1.28$ and $1.86$ for GMFM and LMFM,  respectively, with $(k_1=3,k_2=4)$). Clearly, GMFM has a higher goodness of fit for the data.

To further compare the GMFM and LMFM, we employ a rolling-validation procedure as in  \cite{yu2022projected}. For each year $t$ from
2007 to 2022, we repeatedly use the $n$ (bandwidth) years observations before $t$ to fit the matrix-variate factor model and
estimate the two loading matrices. The loadings are then used to estimate the factors and corresponding residuals of the $4$ quarters in the current year. Specifically, let $ {Y}_t^i$ and $\widehat{ {Y}}_t^i$
be the observed and estimated performance matrix of quarter $i$ in year $t$. Define
\begin{align*}
    \text{MSE}_t = \frac{1}{4\times 12\times 22}\sum_{i=1}^4\left\Vert \widehat{ {Y}}_t^i- {Y}_t^i\right\Vert_F^2, \ \mbox{and} \ \rho_t = \frac{\sum_{i=1}^4\left\Vert \widehat{ {Y}}_t^i- {Y}_t^i\right\Vert_F^2}{\sum_{i=1}^4\left\Vert { {Y}}_t^i-\overline{ {Y}}_t\right\Vert_F^2},
\end{align*}
as the mean squared error (MSE) and unexplained proportion of total variances, respectively. Let $\overline{\text{MSE}}$ and $\bar{\rho}$ be the
mean MSE and the mean unexplained proportion of total variances, respectively. For LMFM, the $\widehat{{Y}}_t^i$'s are reconstructed by the projected estimator (PE) in \cite{yu2022projected} and the $\alpha$-PCA method with $\alpha = 0$ ($\alpha$-PCA) in Chen et al. (2020a).
 \begin{table}[htbp]
\caption{Rolling validation for the operating performance of the 12 listed high-tech manufacturing companies by different models.}
\label{Tab:realrolling}
\begin{center}
\resizebox{\linewidth}{!}{
\begin{tabular}{cccccccccc}
\hline
&&\multicolumn{4}{c}{${\overline{MSE}}$}&\multicolumn{4}{c}{$\bar{\rho}$}\\
\cmidrule(r){3-6}\cmidrule(r){7-10}
n&$k$ &LMFM  &LMFM &GMFM &GVFM  &LMFM  &LMFM &GMFM&GVFM \\
& &($\alpha$-PCA) &(PE) & & &($\alpha$-PCA) &(PE) &&\\ \hline
$5$&1& 0.97 &0.99 &$\mathbf{0.92}$ &0.93 &1.20&1.21&$\mathbf{1.12}$&1.13\\
5&2& 0.90&0.91&$\mathbf{0.85}$&0.91&1.11&1.10&$\mathbf{1.02}$&1.12\\
5&3& 0.81 &0.80 &$\mathbf{0.78}$ &0.88 &0.99 &0.96 &$\mathbf{0.95}$ &1.07\\
10&1&  0.96 &0.96 &$\mathbf{0.91}$ &$\mathbf{0.91}$ &1.21&1.19&$\mathbf{1.12}$&1.13\\
10&2& 0.88 &0.87 &$\mathbf{0.83}$ &0.89 &1.10 &1.08 &$\mathbf{1.02}$ &1.14\\
10&3& 0.79 &0.77 &$\mathbf{0.75}$ &0.88 &0.97 &$0.95$ &$\mathbf{0.93}$ &1.08\\
\hline
\end{tabular}}
\end{center}
\end{table}

Table \ref{Tab:realrolling} presents the average values of MSE and $\rho$ obtained from various estimation methods: LMFM ($\alpha$-PCA), LMFM (PE), GMFM and the generalized vector factor model (GVFM) with $k^2$ factor numbers by \cite{Liu2023Gene}. A comprehensive comparison is made across different combinations of bandwidth ($n$) and the number of factors ($k_1 = k_2 = k$).
From Table \ref{Tab:realrolling}, our proposed GMFM outperforms other methods across the board. Specifically, when k=1, GMFM and GVFM display comparable MSEs. Nevertheless, as k rises, the MSE of GMFM notably decreases, whereas the decline for GVFM is gradual. Moreover, the two estimation methods for LMFM demonstrate comparable behavior, both worse than GMFM.

\section{Conclusion}\label{CD}
This paper provides a general theory for matrix factor analysis of a high-dimensional nonlinear model. Given certain regularity conditions, we have established the consistency and convergence rates of the estimated factors and loadings. Meanwhile, the asymptotic normality of the estimated loadings is derived. Furthermore, we develop a criterion based on a penalized loss to consistently estimate the number of row factors and column factors within the framework of the GMFM. Computationally, We have proposed two algorithms to compute the factors and loadings from the GMFM. To justify our theory, we conduct extensive simulation studies and applied GMFM to the analysis of a business performance dataset.

\section*{Supplementary Material}
The Supplementary Material contains the detailed technical proof of the main theorems, as well as some technical lemmas that are of their own interests, in addition to details of the data set.

\bibliographystyle{elsarticle-harv}
\bibliography{ref}

\begin{thebibliography}{32}
\expandafter\ifx\csname natexlab\endcsname\relax\def\natexlab#1{#1}\fi
\providecommand{\url}[1]{\texttt{#1}}
\providecommand{\href}[2]{#2}
\providecommand{\path}[1]{#1}
\providecommand{\DOIprefix}{doi:}
\providecommand{\ArXivprefix}{arXiv:}
\providecommand{\URLprefix}{URL: }
\providecommand{\Pubmedprefix}{pmid:}
\providecommand{\doi}[1]{\href{http://dx.doi.org/#1}{\path{#1}}}
\providecommand{\Pubmed}[1]{\href{pmid:#1}{\path{#1}}}
\providecommand{\bibinfo}[2]{#2}
\ifx\xfnm\relax \def\xfnm[#1]{\unskip,\space#1}\fi
\bibitem[{Bai and Li(2012)}]{bai2012statistical}
\bibinfo{author}{Bai, J.}, \bibinfo{author}{Li, K.}, \bibinfo{year}{2012}.
\newblock \bibinfo{title}{Statistical analysis of factor models of high dimension}.
\newblock \bibinfo{journal}{The Annals of Statistics} \bibinfo{volume}{40}, \bibinfo{pages}{436--465}.
\bibitem[{Bai and Ng(2002)}]{Bai2002De}
\bibinfo{author}{Bai, J.}, \bibinfo{author}{Ng, S.}, \bibinfo{year}{2002}.
\newblock \bibinfo{title}{Determining the number of factors in approximate factor models}.
\newblock \bibinfo{journal}{Econometrica} \bibinfo{volume}{70}, \bibinfo{pages}{191--221}.
\bibitem[{Barigozzi et~al.(2022)Barigozzi, He, Li and Trapani}]{Barigozzi2022StatisticalIF}
\bibinfo{author}{Barigozzi, M.}, \bibinfo{author}{He, Y.}, \bibinfo{author}{Li, L.}, \bibinfo{author}{Trapani, L.}, \bibinfo{year}{2022}.
\newblock \bibinfo{title}{Statistical inference for large-dimensional tensor factor model by iterative projections}.
\newblock \bibinfo{journal}{arXiv:2206.09800} .
\bibitem[{Barigozzi and Trapani(2022)}]{barigozzi2022testing}
\bibinfo{author}{Barigozzi, M.}, \bibinfo{author}{Trapani, L.}, \bibinfo{year}{2022}.
\newblock \bibinfo{title}{Testing for common trends in nonstationary large datasets}.
\newblock \bibinfo{journal}{Journal of Business \& Economic Statistics} \bibinfo{volume}{40}, \bibinfo{pages}{1107--1122}.
\bibitem[{Chang et~al.(2023)Chang, He, Yang and Yao}]{chang2023modelling}
\bibinfo{author}{Chang, J.}, \bibinfo{author}{He, J.}, \bibinfo{author}{Yang, L.}, \bibinfo{author}{Yao, Q.}, \bibinfo{year}{2023}.
\newblock \bibinfo{title}{Modelling matrix time series via a tensor cp-decomposition}.
\newblock \bibinfo{journal}{Journal of the Royal Statistical Society Series B: Statistical Methodology} \bibinfo{volume}{85}, \bibinfo{pages}{127--148}.
\bibitem[{Chen et~al.(2024)Chen, Fan and Zhu}]{chen2024factor}
\bibinfo{author}{Chen, E.}, \bibinfo{author}{Fan, J.}, \bibinfo{author}{Zhu, X.}, \bibinfo{year}{2024}.
\newblock \bibinfo{title}{Factor augmented matrix regression}.
\newblock \bibinfo{journal}{arXiv:2405.17744} .
\bibitem[{Chen and Fan(2023)}]{Chen2023Statistical}
\bibinfo{author}{Chen, E.Y.}, \bibinfo{author}{Fan, J.}, \bibinfo{year}{2023}.
\newblock \bibinfo{title}{Statistical inference for high-dimensional matrix-variate factor models}.
\newblock \bibinfo{journal}{Journal of the American Statistical Association} \bibinfo{volume}{118}, \bibinfo{pages}{1038--1055}.
\bibitem[{Chen et~al.(2021a)Chen, Fern¨¢ndez-Val and Weidner}]{CHENnonlinear}
\bibinfo{author}{Chen, M.}, \bibinfo{author}{Fern¨¢ndez-Val, I.}, \bibinfo{author}{Weidner, M.}, \bibinfo{year}{2021}a.
\newblock \bibinfo{title}{Nonlinear factor models for network and panel data}.
\newblock \bibinfo{journal}{Journal of Econometrics} \bibinfo{volume}{220}, \bibinfo{pages}{296--324}.
\bibitem[{Chen et~al.(2021b)Chen, Xiao and Yang}]{chen2021auto}
\bibinfo{author}{Chen, R.}, \bibinfo{author}{Xiao, H.}, \bibinfo{author}{Yang, D.}, \bibinfo{year}{2021}b.
\newblock \bibinfo{title}{Autoregressive models for matrix-valued time series}.
\newblock \bibinfo{journal}{Journal of Econometrics} \bibinfo{volume}{222}, \bibinfo{pages}{539--560}.
\bibitem[{Chen et~al.(2022)Chen, Yang and Zhang}]{chen2022factor}
\bibinfo{author}{Chen, R.}, \bibinfo{author}{Yang, D.}, \bibinfo{author}{Zhang, C.H.}, \bibinfo{year}{2022}.
\newblock \bibinfo{title}{Factor models for high-dimensional tensor time series}.
\newblock \bibinfo{journal}{Journal of the American Statistical Association} \bibinfo{volume}{117}, \bibinfo{pages}{94--116}.
\bibitem[{Doz et~al.(2012)Doz, Giannone and Reichlin}]{Doz2012}
\bibinfo{author}{Doz, C.}, \bibinfo{author}{Giannone, D.}, \bibinfo{author}{Reichlin, L.}, \bibinfo{year}{2012}.
\newblock \bibinfo{title}{A quasi-maximum likelihood approach for large, approximate dynamic factor models}.
\newblock \bibinfo{journal}{The Review of Economics and Statistics} \bibinfo{volume}{94}, \bibinfo{pages}{1014--1024}.
\bibitem[{Fan et~al.(2013)Fan, Liao and Mincheva}]{fan2013large}
\bibinfo{author}{Fan, J.}, \bibinfo{author}{Liao, Y.}, \bibinfo{author}{Mincheva, M.}, \bibinfo{year}{2013}.
\newblock \bibinfo{title}{Large covariance estimation by thresholding principal orthogonal complements}.
\newblock \bibinfo{journal}{Journal of the Royal Statistical Society Series B: Statistical Methodology} \bibinfo{volume}{75}, \bibinfo{pages}{603--680}.
\bibitem[{Goyal et~al.(2008)Goyal, P¨¦rignon and Villa}]{GOYAL2008252}
\bibinfo{author}{Goyal, A.}, \bibinfo{author}{P¨¦rignon, C.}, \bibinfo{author}{Villa, C.}, \bibinfo{year}{2008}.
\newblock \bibinfo{title}{How common are common return factors across the nyse and nasdaq?}
\newblock \bibinfo{journal}{Journal of Financial Economics} \bibinfo{volume}{90}, \bibinfo{pages}{252--271}.
\bibitem[{He et~al.(2022)He, Wang, Yu, Zhou and Zhou}]{he2024matrixk}
\bibinfo{author}{He, Y.}, \bibinfo{author}{Wang, Y.}, \bibinfo{author}{Yu, L.}, \bibinfo{author}{Zhou, W.}, \bibinfo{author}{Zhou, W.}, \bibinfo{year}{2022}.
\newblock \bibinfo{title}{Matrix kendall's tau in high-dimensions: A robust statistic for matrix factor model}.
\newblock \bibinfo{journal}{arXiv: 2207.09633} .
\bibitem[{He et~al.(2023)He, Zhao and Zhou}]{he2023Iterative}
\bibinfo{author}{He, Y.}, \bibinfo{author}{Zhao, R.}, \bibinfo{author}{Zhou, W.}, \bibinfo{year}{2023}.
\newblock \bibinfo{title}{Iterative alternating least square estimation for large-dimensional matrix factor model}.
\newblock \bibinfo{journal}{arXiv:2301.00360} .
\bibitem[{Jennrich(1969)}]{Jennrich1969}
\bibinfo{author}{Jennrich, R.I.}, \bibinfo{year}{1969}.
\newblock \bibinfo{title}{Asymptotic properties of non-linear least squares estimators}.
\newblock \bibinfo{journal}{The Annals of Mathematical Statistics} \bibinfo{volume}{40}, \bibinfo{pages}{633--643}.
\bibitem[{Jing et~al.(2021)Jing, Li, Lyu and Xia}]{Jing2021Community}
\bibinfo{author}{Jing, B.}, \bibinfo{author}{Li, T.}, \bibinfo{author}{Lyu, Z.}, \bibinfo{author}{Xia, D.}, \bibinfo{year}{2021}.
\newblock \bibinfo{title}{Community detection on mixture multilayer networks via regularized tensor decomposition}.
\newblock \bibinfo{journal}{Annals of Statistics} \bibinfo{volume}{49}, \bibinfo{pages}{3181--3205}.
\bibitem[{Jing et~al.(2022)Jing, Li, Ying and Yu}]{Jing2022Community}
\bibinfo{author}{Jing, B.}, \bibinfo{author}{Li, T.}, \bibinfo{author}{Ying, N.}, \bibinfo{author}{Yu, X.}, \bibinfo{year}{2022}.
\newblock \bibinfo{title}{Community detection in sparse networks using the symmetrized laplacian inverse matrix (slim)}.
\newblock \bibinfo{journal}{Statistica Sinica} \bibinfo{volume}{32}, \bibinfo{pages}{1--22}.
\bibitem[{Kong(2017)}]{kong2017number}
\bibinfo{author}{Kong, X.}, \bibinfo{year}{2017}.
\newblock \bibinfo{title}{On the number of common factors with high-frequency data}.
\newblock \bibinfo{journal}{Biometrika} \bibinfo{volume}{104}, \bibinfo{pages}{397--410}.
\bibitem[{Liu et~al.(2023)Liu, Lin, Zheng and Liu}]{Liu2023Gene}
\bibinfo{author}{Liu, W.}, \bibinfo{author}{Lin, H.}, \bibinfo{author}{Zheng, S.}, \bibinfo{author}{Liu, J.}, \bibinfo{year}{2023}.
\newblock \bibinfo{title}{Generalized factor model for ultra-high dimensional correlated variables with mixed types}.
\newblock \bibinfo{journal}{Journal of the American Statistical Association} \bibinfo{volume}{118}, \bibinfo{pages}{1385--1401}.
\bibitem[{Mao et~al.(2019)Mao, Chen and Wong}]{mao2019matrix}
\bibinfo{author}{Mao, X.}, \bibinfo{author}{Chen, S.}, \bibinfo{author}{Wong, R.K.W.}, \bibinfo{year}{2019}.
\newblock \bibinfo{title}{Matrix completion with covariate information}.
\newblock \bibinfo{journal}{Journal of the American Statistical Association} \bibinfo{volume}{114}, \bibinfo{pages}{198--210}.
\bibitem[{Mao et~al.(2021)Mao, Wong and Chen}]{mao2021Matrix}
\bibinfo{author}{Mao, X.}, \bibinfo{author}{Wong, R.K.W.}, \bibinfo{author}{Chen, S.}, \bibinfo{year}{2021}.
\newblock \bibinfo{title}{Matrix completion under low-rank missing mechanism}.
\newblock \bibinfo{journal}{Statistica Sinica} \bibinfo{volume}{31}, \bibinfo{pages}{2005--2030}.
\bibitem[{Newey and McFadden(1986)}]{Newey1986LargeSE}
\bibinfo{author}{Newey, W.}, \bibinfo{author}{McFadden, D.}, \bibinfo{year}{1986}.
\newblock \bibinfo{title}{Large sample estimation and hypothesis testing}.
\newblock \bibinfo{journal}{Handbook of Econometrics} \bibinfo{volume}{4}, \bibinfo{pages}{2111--2245}.
\bibitem[{Pelger(2019)}]{pelger2019large}
\bibinfo{author}{Pelger, M.}, \bibinfo{year}{2019}.
\newblock \bibinfo{title}{Large-dimensional factor modeling based on high-frequency observations}.
\newblock \bibinfo{journal}{Journal of Econometrics} \bibinfo{volume}{208}, \bibinfo{pages}{23--42}.
\bibitem[{Trapani(2018)}]{Trapani2008Aran}
\bibinfo{author}{Trapani, L.}, \bibinfo{year}{2018}.
\newblock \bibinfo{title}{A randomized sequential procedure to determine the number of factors}.
\newblock \bibinfo{journal}{Journal of the American Statistical Association} \bibinfo{volume}{113}, \bibinfo{pages}{1341--1349}.
\bibitem[{Wang et~al.(2019)Wang, Liu and Chen}]{wang2019factor}
\bibinfo{author}{Wang, D.}, \bibinfo{author}{Liu, X.}, \bibinfo{author}{Chen, R.}, \bibinfo{year}{2019}.
\newblock \bibinfo{title}{Factor models for matrix-valued high-dimensional time series}.
\newblock \bibinfo{journal}{Journal of Econometrics} \bibinfo{volume}{208}, \bibinfo{pages}{231--248}.
\bibitem[{Wang(2022)}]{WANG2022180}
\bibinfo{author}{Wang, F.}, \bibinfo{year}{2022}.
\newblock \bibinfo{title}{Maximum likelihood estimation and inference for high dimensional generalized factor models with application to factor-augmented regressions}.
\newblock \bibinfo{journal}{Journal of Econometrics} \bibinfo{volume}{229}, \bibinfo{pages}{180--200}.
\bibitem[{Wu(1981)}]{Wu1981}
\bibinfo{author}{Wu, C.F.}, \bibinfo{year}{1981}.
\newblock \bibinfo{title}{Asymptotic theory of nonlinear least squares estimation}.
\newblock \bibinfo{journal}{The Annals of Statistics} \bibinfo{volume}{9}, \bibinfo{pages}{501--513}.
\bibitem[{Yu et~al.(2022)Yu, He, Kong and Zhang}]{yu2022projected}
\bibinfo{author}{Yu, L.}, \bibinfo{author}{He, Y.}, \bibinfo{author}{Kong, X.}, \bibinfo{author}{Zhang, X.}, \bibinfo{year}{2022}.
\newblock \bibinfo{title}{Projected estimation for large-dimensional matrix factor models}.
\newblock \bibinfo{journal}{Journal of Econometrics} \bibinfo{volume}{229}, \bibinfo{pages}{201--217}.
\bibitem[{Yuan et~al.(2023)Yuan, Gao, He, Huang and Guo}]{yuan2023two}
\bibinfo{author}{Yuan, C.}, \bibinfo{author}{Gao, Z.}, \bibinfo{author}{He, X.}, \bibinfo{author}{Huang, W.}, \bibinfo{author}{Guo, J.}, \bibinfo{year}{2023}.
\newblock \bibinfo{title}{Two-way dynamic factor models for high-dimensional matrix-valued time series}.
\newblock \bibinfo{journal}{Journal of the Royal Statistical Society Series B: Statistical Methodology} \bibinfo{volume}{85}, \bibinfo{pages}{1517--1537}.
\bibitem[{Zhang et~al.(2024)Zhang, Liu, Guo, Yuen and Welsh}]{zhang2024mod}
\bibinfo{author}{Zhang, X.}, \bibinfo{author}{Liu, C.C.}, \bibinfo{author}{Guo, J.}, \bibinfo{author}{Yuen, K.C.}, \bibinfo{author}{Welsh, A.H.}, \bibinfo{year}{2024}.
\newblock \bibinfo{title}{Modeling and learning on high-dimensional matrix-variate sequences}.
\newblock \bibinfo{journal}{Journal of the American Statistical Association} \bibinfo{volume}{0}, \bibinfo{pages}{1--16}.
\newblock \URLprefix \url{https://doi.org/10.1080/01621459.2024.2344687}.
\bibitem[{Zhou et~al.(2024)Zhou, Pan, Zhang and Wang}]{zhou2024attribute}
\bibinfo{author}{Zhou, T.}, \bibinfo{author}{Pan, R.}, \bibinfo{author}{Zhang, J.}, \bibinfo{author}{Wang, H.}, \bibinfo{year}{2024}.
\newblock \bibinfo{title}{An attribute-based node2vec model for dynamic community detection on co-authorship network}.
\newblock \bibinfo{journal}{Computational Statistics} , \bibinfo{pages}{1--28}.

\end{thebibliography}
\end{document}